\documentclass[fleqn,usenatbib]{mnras}

\usepackage{newtxtext,newtxmath}

\usepackage[T1]{fontenc}
\usepackage{ae,aecompl}
\usepackage{comment}
\usepackage{multirow}

\usepackage{graphicx}	% Including figure files
\usepackage{amsmath}	% Advanced maths commands
\usepackage{amssymb}	% Extra maths symbols

\usepackage{xcolor}  % for highlighting
\usepackage{ulem}  % for editing

\usepackage{threeparttable}

\title[Chemical abundances in old halo star candidates]{Origin of metals in old Milky Way halo stars based on GALAH and Gaia} 

\author[Miho N. Ishigaki et al.]{Miho N. Ishigaki,$^{1,2,4}$\thanks{E-mail: miho.ishigaki@nao.ac.jp (MNI)}
Tilman Hartwig,$^{3,4,5}$
Yuta Tarumi,$^{5}$
Shing-Chi Leung,$^{4,6}$ 
\newauthor
Nozomu Tominaga,$^{1,4,7}$
Chiaki Kobayashi,$^{8,4}$
Mattis Magg,$^{9, 10}$
\newauthor
Aurora Simionescu,$^{4,11,12}$
Ken'ichi Nomoto$^{4}$
\\
$^{1}$ National Astronomical Observatory of Japan, 2-21-1 Osawa, Mitaka, Tokyo 181-8588, Japan \\
$^{2}$ Astronomical Institute, Tohoku University, 6-3, Aramaki, Aoba-ku, Sendai, Miyagi, 980-8578, Japan \\ 
$^{3}$Institute for Physics of Intelligence, School of Science, The University of Tokyo, Bunkyo, Tokyo 113-0033, Japan\\
$^{4}$Kavli Institute for the Physics and Mathematics of the Universe (WPI), The University of Tokyo Institutes for Advanced Study, \\ 
The University of Tokyo, 5-1-5
Kashiwanoha, Kashiwa, Chiba 277-8583, Japan\\
$^{5}$Department of Physics, School of Science, The University of Tokyo, Bunkyo, Tokyo 113-0033, Japan\\
$^{6}$ TAPIR, Walter Burke Institute for Theoretical Physics, Mailcode 350-17, \\
California Institute of Technology, Pasadena, CA 91125, USA\\
$^{7}$ Department of Physics, Faculty of Science and Engineering, Konan University, 8-9-1 Okamoto, Kobe, Hyogo 658-8501, Japan\\
$^{8}$Centre for Astrophysics Research, Department of Physics, Astronomy and Mathematics, \\
University of Hertfordshire, Hatfield AL10 9AB, UK\\
$^{9}$ Institut f\"{u}r Theoretische Astrophysik, Zentrum f\"{u}r Astronomie, Universit\"{a}t Heidelberg, D-69120 Heidelberg, Germany\\
$^{10}$ International Max Planck Research School for Astronomy and Cosmic Physics at the University of Heidelberg (IMPRS-HD), \\
D-69117 Heidelberg, Germany \\
$^{11}$ SRON Netherlands Institute for Space Research, Sorbonnelaan 2, 3584 CA Utrecht, The Netherlands \\
$^{12}$ Leiden Observatory, Leiden University, PO Box 9513, 2300 RA Leiden, The Netherlands \\
}

\date{Accepted XXX. Received YYY; in original form ZZZ}

\pubyear{2015}
\begin{document}
\label{firstpage}
\pagerange{\pageref{firstpage}--\pageref{lastpage}}

\maketitle

\begin{abstract}

Stellar and supernova nucleosynthesis in the 
first few billion years of the cosmic history 
have set the scene for early structure formation in the Universe, 
while little is known about their nature.  
Making use of stellar physical parameters measured by GALAH 
Data Release 3 with accurate astrometry from the Gaia EDR3, we have selected $\sim 100$ old
main-sequence turn-off stars (ages $\gtrsim 12$\,Gyrs) with kinematics compatible with the Milky Way stellar halo population in the Solar neighborhood. 
Detailed 
homogeneous elemental abundance estimates by GALAH DR3 are compared with supernova yield models of Pop~III (zero-metal) 
core-collapse supernovae 
(CCSNe), normal (non-zero-metal) CCSNe, and Type Ia supernovae 
(SN~Ia) to examine which of the 
individual yields or their combinations best reproduce 
the observed elemental abundance patterns for each 
of the old halo stars (``OHS''). 
We find that the observed abundances in the OHS with [Fe/H]$>-1.5$ 
are best explained by contributions from both CCSNe and SN~Ia, 
where the fraction of SN~Ia among all the metal-enriching SNe 
is up to 10--20 \% for stars with high [Mg/Fe] ratios and 
up to 20--27 \% for stars with low [Mg/Fe] ratios, depending on the assumption about the relative fraction of near-Chandrasekhar-mass SNe~Ia progenitors.
The results suggest that, in the progenitor 
systems of the OHS with [Fe/H]$>-1.5$, $\sim$ 50--60\% of Fe mass originated from normal CCSNe at the earliest phases of 
the Milky Way formation. These results provide an insight into 
the birth environments of the oldest stars in the Galactic halo.

\end{abstract}


%%%%%%%%%%%%%%%%%%%%%%%%%%%%%%%%%%%%%%%%%%%%%%%%%%

%%%%%%%%%%%%%%%%% BODY OF PAPER %%%%%%%%%%%%%%%%%%

\section{Introduction}

Production of elements by stars and supernovae in the 
first few billion years of the cosmic history 
had impacted the environment of star formation 
in the early Universe 
 \citep{bromm04,greif10,bromm11,karlsson13}.
After the Big Bang nucleosynthesis, the first 
stars, so called Population III (Pop~III) stars, were
formed as a result of condensation of 
primordial, pure H and He gas in cosmological mini-halos driven 
by cooling via molecular hydrogen \citep[e.g.,][]{bromm04}. Depending on the initial stellar masses, Pop~III 
stars undergo supernova explosions and have 
enriched the primordial gas with 
metals for the first time in the cosmic history \citep{umeda02,heger02, heger10,limongi12, nomoto13, ishigaki18}. 

Once a critical metal or dust abundance is reached,  
gas cooling becomes more efficient,
which leads to the formation of stars with masses more 
typical of the present-day Universe \citep[e.g.,][]{bromm11,chiaki14}.
The stars with masses in the range $\sim 10-40~M_\odot$ 
mainly synthesize elements from carbon to silicon 
through their hydrostatic burning and eject the 
nucleosynthetic products by 
core-collapse supernovae (CCSNe) 
\citep[e.g.,][]{woosley95,thielemann96,nomoto13}. 
These normal CCSNe (progenitors with non-zero metal contents) further 
chemically enrich the interstellar medium.  
Stars with masses below $\sim 8~M_\odot$ evolve into a white dwarf, which is 
a potential progenitor of a Type Ia supernova (SN~Ia) \citep[e.g.,][]{umeda99,kobayashi20a}. 
 The SNe~Ia are mainly responsible for 
production of Fe-group elements, such that 
60 \% of Fe 
in the Solar system material is originated from SN~Ia (\citealt{kobayashi20b}, see also \citealt{tsujimoto95}).

The chemical enrichment processes by these nucleosynthetic 
channels in the early galactic environments are still elusive. The physical properties of Pop~III stars are one of the biggest uncertainties regarding the early cosmic chemical evolution. 
Cosmological simulations predict several orders of magnitudes difference in characteristic masses of the Pop~III stars 
ranging from more than $\sim$100$~M_\odot$ \citep{bromm02,oshea07,yoshida06}, a few tens of $M_\odot$  \citep{stacy10,stacy12,hosokawa11,sharda19}, 
down to less than a few $M_\odot$ \citep{clark11,greif11,stacy14}. 
A few orders of magnitudes 
spread in the predicted Pop~III 
masses and their multiplicity have also been predicted \citep{hirano14,hirano15,susa14,susa19,sugimura20}. 
In addition to the masses, presence of stellar rotation could also impact the 
metal yields as well as the overall evolution of the progenitor stars
\citep{meynet06,hirschi07,joggerst10,takahashi14,limongi18,choplin19}. 

In addition to the Pop~III properties, it remains 
unclear how the ejected metals from Pop~III supernovae are 
transported in the inter galactic medium (IGM) and mixed with primordial 
gas, from which the next generation, namely, the first metal-enriched stars formed \citep[e.g.,][]{smith09,ritter12,chiaki18,tarumi20}. Such 
mechanisms can depend on multiplicity of 
Pop~III stars \citep{hartwig18,hartwig19a} or
 presence of radiative feedback from 
the Pop~III stars \citep{greif10,jeon14,cooke14,chiaki18}.

Physical mechanisms of CCSN explosions
of massive stars in general 
are crucial for the final yields, while they are not well 
understood \citep[e.g.,][]{janka12}. The explosive nucleosynthesis, amount of fallback \citep[e.g.,][]{woosley95,thielemann96,zhang08}, mixing 
in ejecta \citep[e.g.,][]{joggerst09}, and departure from spherical symmetry \citep[e.g.,][]{tominaga09,ezzeddine19}
can all affect the nucleosynthesis products that ultimately contribute to the chemical enrichment. 

Finally, progenitor systems of SNe~Ia 
have not been identified. 
 SN~Ia is 
 a thermonuclear explosion of a white dwarf 
 of carbon-oxygen composition. A minimum possible 
 time scale is thus determined 
 by an age of the most massive white dwarf progenitors, 
 which corresponds to 
 a few tens of Myrs for $\sim 8~M_\odot$ stars. 
 Whether a white dwarf is able to explode as a SN~Ia 
 depends on properties of the host 
 binary system, which determines mass accretion from 
 the companion star \citep[Single-Degenerate or SD scenario][]{nomoto82}
 or on the possibility of a merger of two white dwarfs within a reasonable 
 timescale \citep[Double-Degenerate or DD scenario, e.g.,][]{hillebrandt00}. 
These scenarios are observationally tested, for example,  
through characteristic delay time for SNe~Ia relative to a major 
star formation episode. The characteristic 
delay time or the delay-time distribution has been addressed 
at various environments and redshifts, while 
a clear consensus about the dominant SN~Ia channel has 
not been obtained \citep[e.g.,][]{hopkins06,totani08,hachisu08,maoz14}.
  In both the SD and DD scenarios, 
  nucleosynthetic yields of 
 SN~Ia are mainly determined by 
 physical conditions such as matter density 
 at which thermonuclear burning is ignited, which 
 depends on masses of white dwarfs at the onset of 
 explosion \citep[e.g.,][]{nomoto84b,thielemann86,sato15,sato16,kobayashi20b}. According to this mass, nucleosynthetic 
 products of SN~Ia are often categorized according to whether 
 the mass of a white dwarf SN~Ia progenitor is close
 to or below the 
 Chandrasekhar white dwarf mass limit, $M_{\rm Ch}\sim 1.4~M_\odot$. 
For some elements, the nucleosynthetic yields of SN~Ia 
are largely different between near-$M_{\rm Ch}$ or 
sub-$M_{\rm Ch}$ white dwarf progenitor models \citep[e.g.,][]{nomoto18,leung18,leung20}.

Since it is not feasible to directly observe
processes of metal enrichment at high redshifts, the only observational 
probe of the metal-enrichment sources in the 
early Universe has been chemical signatures retained in the 
atmosphere of nearby long-lived stars in the Milky Way halo. 
Stars with the lowest Fe abundances such as 
extremely metal-poor (EMP) stars with [Fe/H]$<-3$  
are commonly considered 
to be 
objects retaining 
chemical signatures of the Pop~III nucloesynthesis \citep{beers05,frebel15}. Stars with a wider range of 
[Fe/H] can also be used as an important probe of chemical evolution 
as a function of cosmic time by comparing with chemical 
evolution models \citep[e.g.,][]{kobayashi06,kobayashi20a}.

While the [Fe/H] abundance is frequently used as a 
proxy for stellar ages {\bf for Galactic halo stars}, it has been shown that the age-[Fe/H] relationship depends on the star formation environment. In fact, the Galactic bulge is known to host very old stellar populations, including globular clusters with an age as old as the age of the Universe \citep[e.g.,][and references therein]{barbuy18}. Cosmological simulations that implement chemical evolution models also
predict that the oldest stars in various Galactic environments do not exclusively possess very low Fe abundances \citep[e.g.,][]{salvadori10,starkenburg17,elbadry18,salvadori19}. 
Instead, stellar ages 
are a fundamental property to identify stars 
more likely formed in the early Universe, and thus, 
under 
the influence of chemical enrichment by the earliest generations of 
stars.
The stellar ages, however, are 
extremely difficult to estimate for a single star and are often 
suffer from large systematic uncertainty or dependent on 
assumptions in stellar evolution modeling \citep{soderblom10}.
It has become feasible only recently to obtain stellar ages for a large 
homogeneous sample of stars thanks to large surveys of Galactic 
stellar populations \citep[e.g.,][]{sanders18,sharma20}. 
 The stars with age, kinematics and chemical abundance estimates 
available from these massive surveys motivate us to address the question whether 
there are old but comparatively metal-rich nearby halo stars 
which possess chemical signatures of the metal enrichment by the earliest 
generations of stars. 

In this paper, 
we select candidates of old stars in the Milky Way halo
based on stellar ages and kinematics provided by the most 
up-to-date data releases of the Galactic Archaeology with HERMES (GALAH) survey \citep{desilva15,buder21} 
and the Gaia mission \citep{lindegren20}.  
For the selected stars, we compare observed elemental abundance patterns with sets of yield models for Pop~III CCSNe, normal (non-zero-metal 10--40$~M_\odot$) CCSNe and SNe~Ia. 

This paper is organized as follows. Section \ref{sec:data} describes the 
sample selection method based on the GALAH and Gaia data. 
Section \ref{sec:yield_models} 
describes the yield models we use to fit the observed elemental 
abundances. Section \ref{sec:results} presents results
of fitting the yield models to the observed abundances. Section \ref{sec:discussion} gives 
implications of the results on the metal-enrichment in the early 
Galactic environment. 
Finally, 
we give a summary and our 
conclusions in Section \ref{sec:conclusion}.

\section{Data}
\label{sec:data}

% Boldface start 

In this paper, we base our analysis on 
stellar parameters, elemental abundances, 
and age estimates from GALAH DR3 \citep{buder21}. 
The main catalog of GALAH DR3 provides 
stellar parameters and 
elemental abundances for 
588,571 stars obtained using both astrometric 
data from Gaia DR2 and spectroscopic 
data from the GALAH survey itself. We further 
make use of one of the 
Value-Added Catalogs, which provides age estimates 
(See Section \ref{sec:age_selection}). 
To calculate stellar orbital parameters, 
GALAH DR3 catalog is cross-matched  
with Gaia EDR3 \citep{lindegren20}, 
which provides up-to-date 
astrometric data for most of the 
GALAH DR3 stars. In the following subsections,  
we first describe the adopted quality cuts. 
We then describe the sample selection 
criteria by age and kinematics. 

\subsection{Quality cuts}
\label{sec:quality_cuts} 

\subsubsection{GALAH DR3} 

We applied the following cuts to select stars that are candidates of old stars with 
reliable stellar parameter and age estimates from the 
main and the Value-Added Catalog of GALAH DR3 \citep{buder21}. 

First, we adopt a criterion, \verb|flag_sp=0|, to ensure that stars have reliable astrometry from Gaia DR2 and with no issues raised in 
the data reduction process regarding either 
data quality, peculiar spectral properties (e.g., binary or emission line stars) or 
large $\chi^2$ in the fitting (for the full list, see Table 6 of  \citealt{buder21}).  Next, 
$S/N =$\verb|SNR_C2_IRAF| $>40$ is applied 
to select stars with 
precise stellar parameter estimates. At \verb|SNR_C2_IRAF| $= 40$, \citet{buder21} 
report that the typical precision 
of $T_{\rm eff}$, $\log g$, and [Fe/H] are 
49 K, 0.07 dex, and 0.055 dex, respectively. We further restrict 
our analysis to main-sequence turn-off (MSTO) 
stars by requiring $3.2<\verb|logg|<4.1$ and $5000<\verb|teff|<7000$, which is the same as 
the MSTO selection criteria 
adopted in \cite{sharma21}. 
Finally, we require that more than five 
measurements of [X/Fe], except for Ti or Sc, are available. This ensures that the model parameters
(see Section \ref{sec:yield_models}) can be robustly constrained by 
the measured abundance ratios. 

To summarize, we adopt following 
quality cuts: 

\begin{itemize}
    \item \verb|flag_sp=0| (73\%)
    \item $S/N =$\verb|SNR_C2_IRAF| $>40$ (29\%)
    \item $3.2<\verb|logg|<4.1$ \& $5000<\verb|teff|<7000$ (7\%)
    \item More than five [X/Fe] measurements. (7\%)
\end{itemize}

Numbers in the brackets indicate the percentage of remaining 
stars relative to all stars in the main catalog 
after additionally applying each cut.

\subsubsection{Gaia EDR3}

The selected stars are cross matched with the Gaia EDR3 \citep{lindegren20} to obtain 
accurate astrometric data for the kinematic selection 
in Section \ref{sec:selectkin}. 
We restrict our sample to 
stars with high astrometric quality 
by requiring \verb|ruwe| $<1.4$ as recommended 
in \citet{lindegren18}, where the value of 
\verb|ruwe| (re-normalized unit weight error) 
quantifies the chi-square of the astrometric fit. 
We also exclude stars with negative parallaxes or with \verb|parallax_error|/\verb|parallax|$\ge 0.1$, to exclude objects with unreliable 
distance estimates. After the above quality cuts, $\sim 35,000$ stars ($\sim 6 \%$ of the entire GALAH DR3 catalog) remain.

\subsection{Sample selection}

\subsubsection{Age}
\label{sec:age_selection}

\begin{figure}
    \centering
    \includegraphics[width=8cm]{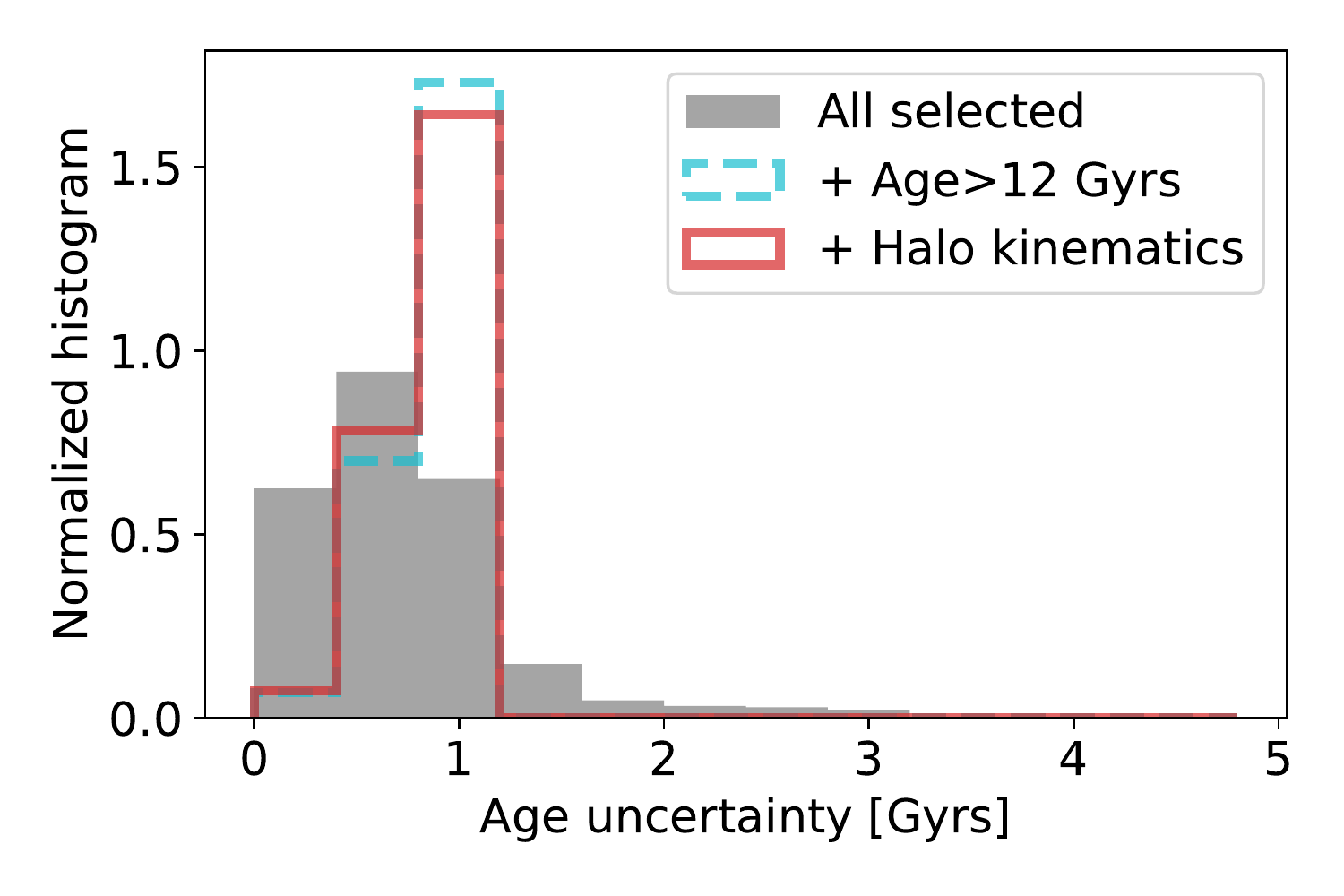}
    \caption{Normalized histograms of the one-sigma age uncertainties from the GALAH DR3 catalog for the sample of stars selected for this study. The gray histogram 
    corresponds to all $\sim 35,000$ stars that satisfy 
    the quality cuts in Section \ref{sec:quality_cuts}. 
    The age-selected sample (Section \ref{sec:age_selection} 
    is shown by the cyan dashed histogram. 
    The stars that satisfy both the age and kinematic 
    criteria (Section \ref{sec:selectkin}) are shown by 
    the red solid histogram. }
    \label{fig:age_uncertainty}
\end{figure}

In order to select old star candidates 
in the GALAH-Gaia 
cross-matched catalog described in the previous section, 
we employ stellar ages from the Value-Added Catalog 
of GALAH DR3 \citep{buder21}. Stellar ages 
in this catalog have 
been estimated by the Bayesian Stellar Parameter Estimation code, BSTEP, 
developed by \citet{sharma18}. The code  
makes use of observed stellar parameters ($T_{\rm eff}$ and $\log g$), 
chemical composition ([Fe/H] and [$\alpha$/Fe]), photometry (2MASS $J$ and $K_{s}$) and astrometry 
 (parallax) to be compared with theoretical isochrone 
models \citep{buder21}. For the 
isochrone models, PARSEC-v1.2S \citep{bressan12} with 121 age grid points spanning $6.6 < \log {\rm age/Gyr} < 10.12$ 
were used \citep{sharma18,buder21}. A flat prior on age across $0< {\rm age/Gyr} < 13.18$ has been adopted \citep{sharma18}. For an estimate of age and its statistical uncertainty in the catalog, a mean and a standard deviation based on 16 and 84 percentiles are reported \citep{buder21}. 
Figure \ref{fig:age_uncertainty} shows histograms of the one-sigma age uncertainties reported in GALAH DR3 for 
the stars that satisfy the quality cuts in Section \ref{sec:quality_cuts} (the gray histogram) and the age-kinematics cut 
described in this and the next section (the cyan and red 
histograms).  
The median value of the age uncertainties for the stars that satisfy the quality cut is 0.7 Gyrs.  We note that the age uncertainties have been obtained assuming the flat age prior, 
which could be a strong prior.
Given that 
typical ages of nearby field Milky Way stellar halo have been estimated to be $> 10$ Gyrs \citep[e.g.,][]{gallart19}, we adopt 
an age cut of $>12$ Gyrs to select 
candidates of very old stars in the 
Solar neighborhood.

\begin{figure*}
    \centering
    \includegraphics[width=17.5cm]{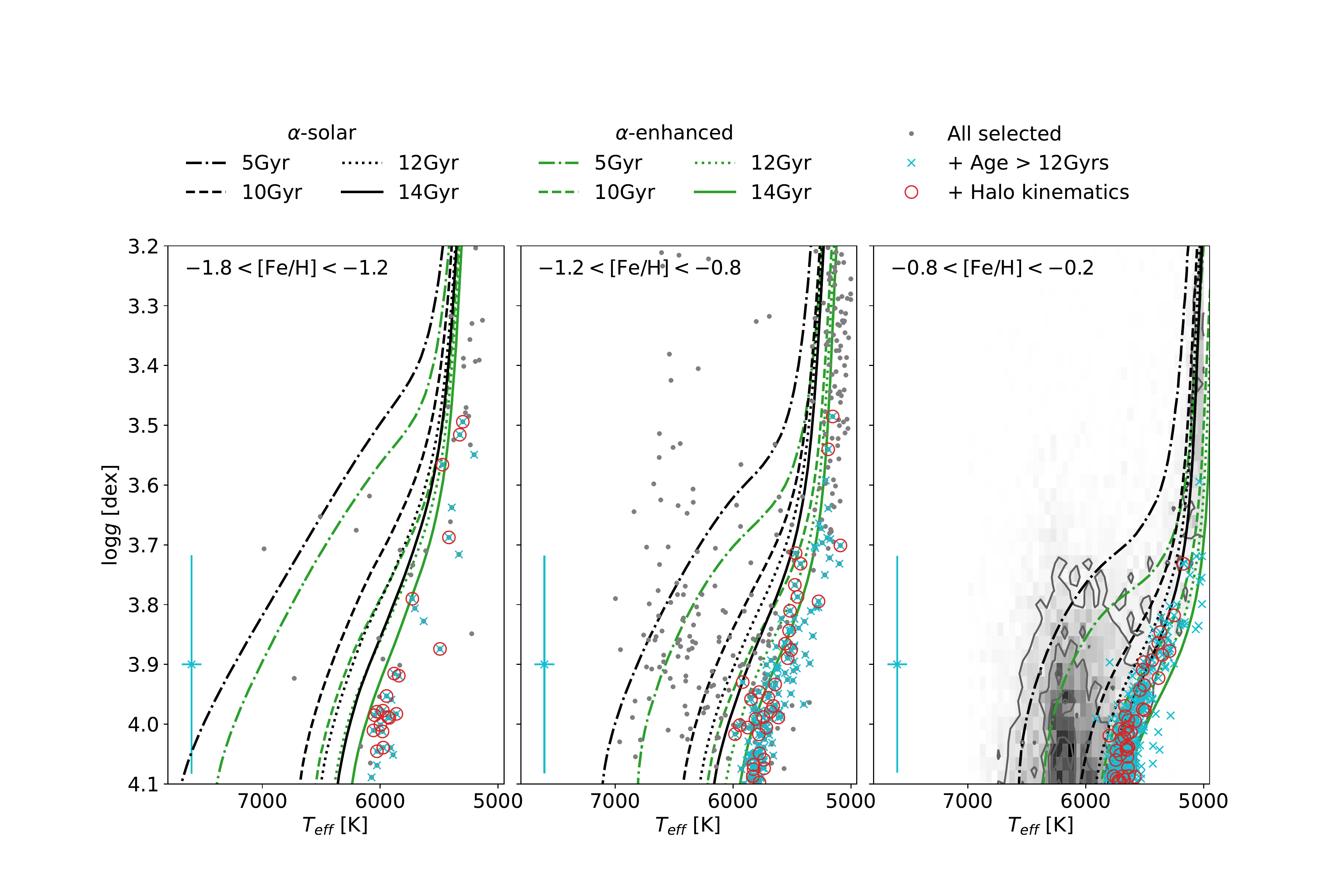}
    \caption{The $T_{\rm eff}-\log g$ diagrams for the sample stars with [Fe/H] metallicities in the ranges $-1.8<$[Fe/H]$\leq -1.2$ ({\it left}), $-1.2<$[Fe/H]$\leq -0.8$ ({\it middle}), 
    and $-0.8<$[Fe/H]$\leq-0.2$ ({\it right}).
    The values of $T_{\rm eff}$ and $\log g$ taken from 
    GALAH DR3 for stars that satisfy the quality cuts in Section \ref{sec:quality_cuts} are plotted 
    by gray dots or the gray contour. 
    Among them, the stars that satisfy the age cut ($>12$ Gyrs) are marked by cyan crosses. 
    Among them, the stars that also satisfy the kinematic criterion (Section \ref{sec:selectkin}) are marked by red circles.
    Typical errors 
    of $T_{\rm eff}$ and $\log g$ from the 
    GALAH DR3 catalog for each [Fe/H] 
    subset is shown at the bottom left. 
     The theoretical isochrones from 
     the Dartmouth Stellar Evolution Database
     \citep{dotter08} with different ages and at [$\alpha$/Fe] = 0.0 are overlaid by dark gray lines, where the oldest age in each [Fe/H] range is shown by a solid line. Corresponding isochrones with an enhanced $\alpha$-element abundance ([$\alpha$/Fe] = 0.4) are shown by green lines.}
    \label{fig:cmd}
\end{figure*}

Figure \ref{fig:cmd} shows $T_{\rm eff}$-$\log g$ diagrams for stars that 
satisfy the age cut (cyan crosses) 
for three different [Fe/H] ranges. A typical 
uncertainty of $T_{\rm eff}$ and $\log g$ 
in the GALAH DR3 catalog 
for the age-selected sample 
is shown at the left bottom in each panel. 
For each [Fe/H] range, 
 theoretical isochrones from the 
Dartmouth Stellar Evolution Database 
\citep{dotter08} with ages 5, 10, 12, and 
14 Gyrs with a scaled-solar $\alpha$-element 
abundance are shown (black dash-dotted, dashed, dotted and solid lines, respectively). 
As can be seen, the age-selected sample 
is mostly cooler than 
the oldest isochrone model. 
Since a sizable fraction of the sample is 
enhanced in [$\alpha$/Fe] in the range 
0.0-0.4 (see Section \ref{sec:chemgroup}), 
Figure \ref{fig:cmd} also shows 
isochrones with an enhanced $\alpha$-element 
abundance of [$\alpha$/Fe] = 0.4. 
It can be seen that, for a given [Fe/H] and an age, 
the $\alpha$-enhanced 
models are cooler than the corresponding 
$\alpha$-solar models and therefore, the 
absolute age estimates could be 
a subject of systematic uncertainties 
depending on [$\alpha$/Fe] abundance ratios 
of individual stars. Because of the 
possible systematic uncertainties in absolute ages, 
our age-selected sample should 
be interpreted as the oldest stars among 
those observed by GALAH DR3 and Gaia EDR3 
in the Solar neighborhood.

\subsubsection{Kinematics} 
\label{sec:selectkin}

In addition to the ages, we further select stars 
with halo-like kinematics based on their parallax and proper motion from Gaia EDR3 and line-of-sight velocities from GALAH DR3.
For this purpose, we require that the total velocity with respect to the Solar velocity to
be greater than 150 km s$^{-1}$ ($|{\bf v_{\rm tot}}-{\bf v}_{\odot}|$ $>150$ km s$^{-1}$). 
With this kinematic cut, 102 stars remain.  Uncertainties in the total velocity are calculated by repeating the velocity calculation 1000 times by adding Gaussian noises to the measured values of
parallax, proper motion and radial velocity  with a standard deviation consistent with the 
corresponding error of each quantity. The mean uncertainty of $|{\bf v_{\rm tot}}-{\bf v}_{\odot}|$ 
obtained for the selected sample is 2.4 km s$^{-1}$ with the standard deviation of 0.7 km s$^{-1}$, which has only a minor effect 
on the kinematic selection.

\begin{figure*}
    \centering
    \includegraphics[width=18cm]{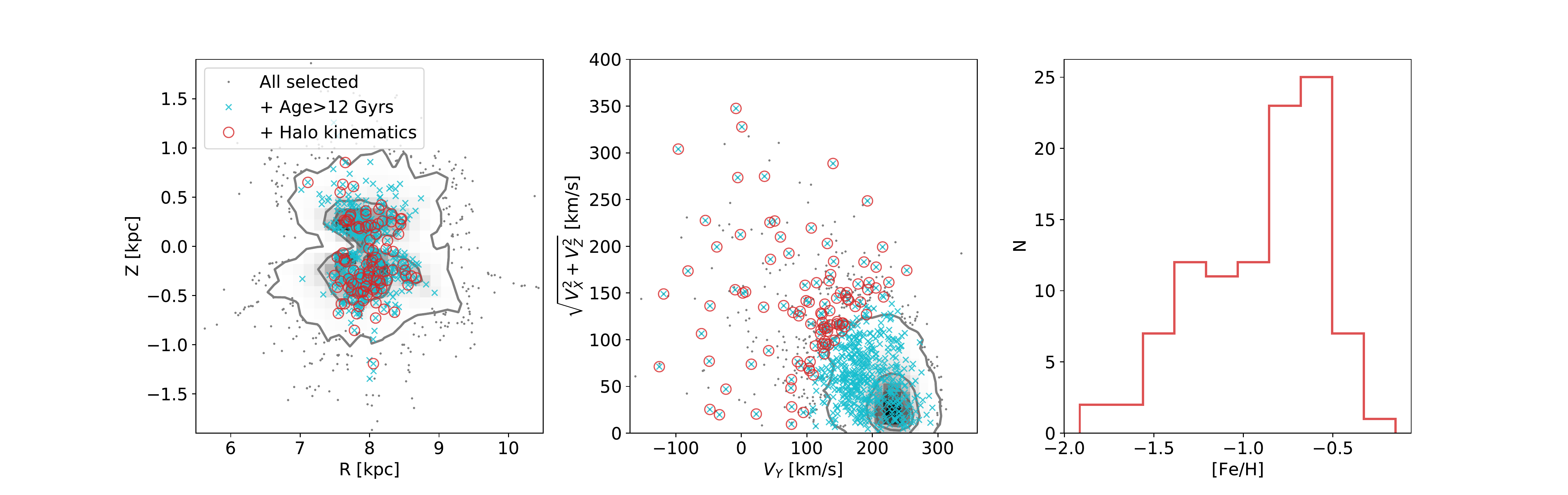}
   \caption{({\it Left}): The spatial distribution of the old halo stars (OHS) selected by the method described in Section \ref{sec:data}. The symbols are the same as in Figure \ref{fig:cmd}. ({\it Middle}): The Toomre diagram for the sample stars. The symbols are the same as in the left plane. ({\it Right}): Metallicity distribution of the OHS sample.}
    \label{fig:sampledist}
\end{figure*}

In this paper, we refer to the selected stars as ``old halo stars (OHS)".  The left panel of Figure \ref{fig:sampledist} shows the spatial 
distribution of the OHS in the Galactic cylindrical coordinates. The majority of 
the OHS selected in this study are located 
within $\sim 1$ kpc above or below the Galactic plane. The middle panel 
of Figure \ref{fig:sampledist} shows 
a Toomre diagram ($v_{Y}$ versus $\sqrt{v_{X}^2 + v_{Z}^2}$) of the OHS 
in the Galactic rest frame. It can be seen that the adopted kinematic cut removes stars with disk-like orbits with 
$V_{Y}> 200$ km s$^{-1}$ as plotted by the gray contours. Finally, the right panel shows the metallicity 
distribution of the OHS, which ranges from $-2$ to $-0.3$.

\begin{figure}
    \centering
    \includegraphics[width=8.5cm]{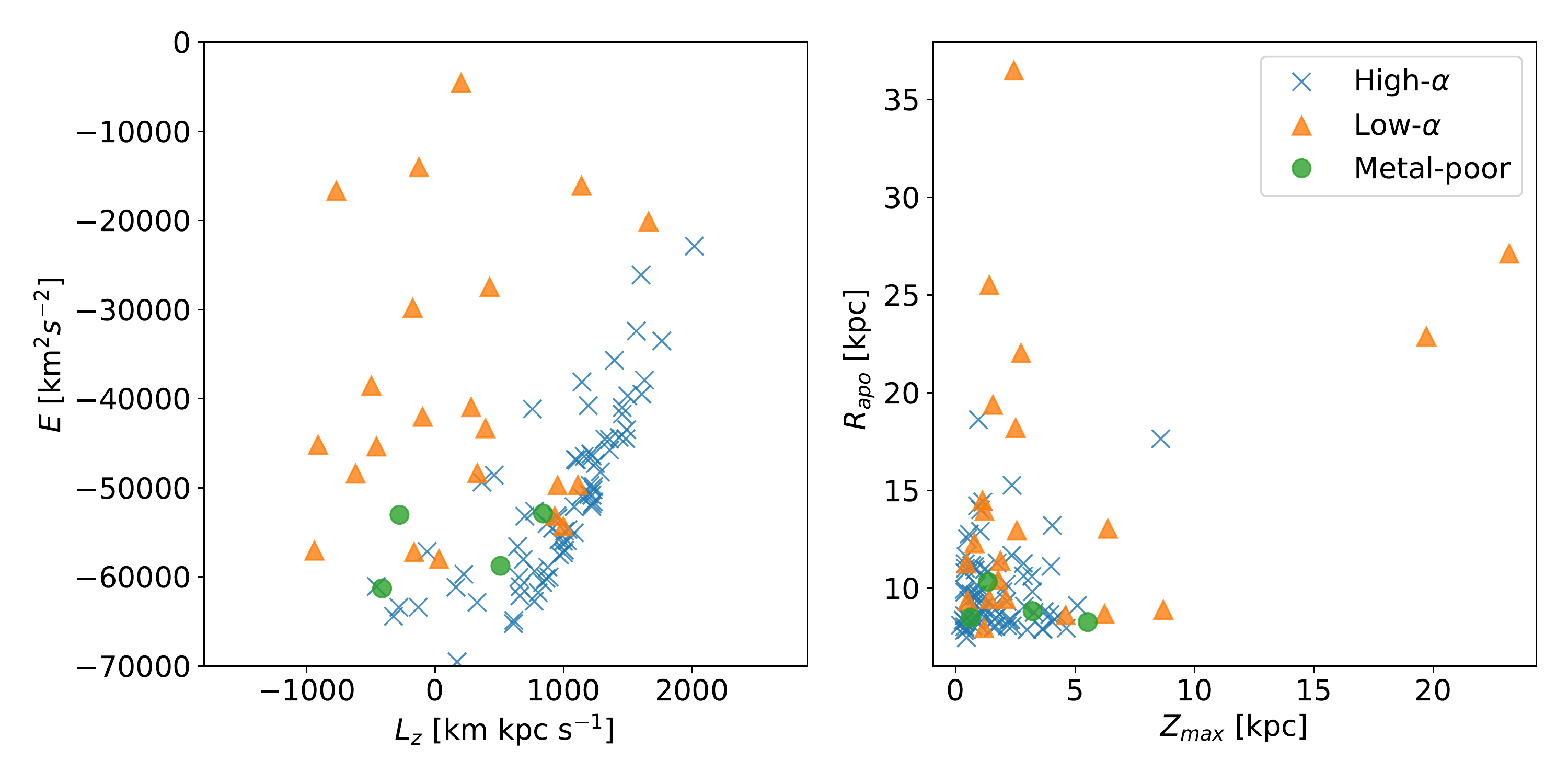}
    \caption{The orbital parameters of the OHS. Different symbols correspond to chemically different subgroups defined in Section \ref{sec:chemgroup} and Figure \ref{fig:xfe_feh}}
    \label{fig:orbit}
\end{figure}

In order to assess whether the selected stars 
kinematically belong to the old Galactic populations, 
orbital parameters of the OHS are 
calculated using the Galpy package \citep{bovy15}\footnote{\url{http://github.com/jobovy/galpy}} assuming the Galactic potential {\verb MWPotential2014 }. The 
resulting orbital energy ($E$) and the $z$ component of the angular 
momentum ($L_{z}$) are shown in the left panel of Figure \ref{fig:orbit}. Apocentric distances ($R_{\rm apo}$) and 
the maximum vertical distances ($Z_{\rm max}$) are shown 
on the right panel of Figure \ref{fig:orbit}. 
Different symbols correspond to subgroups 
defined by chemical abundances as described in 
the next section. 
A large fraction of the OHS exhibits orbital parameters either $R_{\rm apo}>10$ kpc 
or $Z_{\rm max}>1$ kpc, suggesting that they mostly belong to the halo population. 
The OHS sample includes stars with disk-like 
orbits, which could belong to the thick disk population or debris of past accretion events \citep[e.g.,][]{bonaca17,bonaca20,dimatteo19,naidu20,amarante20,montalban21}.

\subsubsection{Chemical abundance subgroups} 
\label{sec:chemgroup}

\begin{figure*}
    \centering
    \includegraphics[width=18cm, trim = 70 70 70 70, clip]{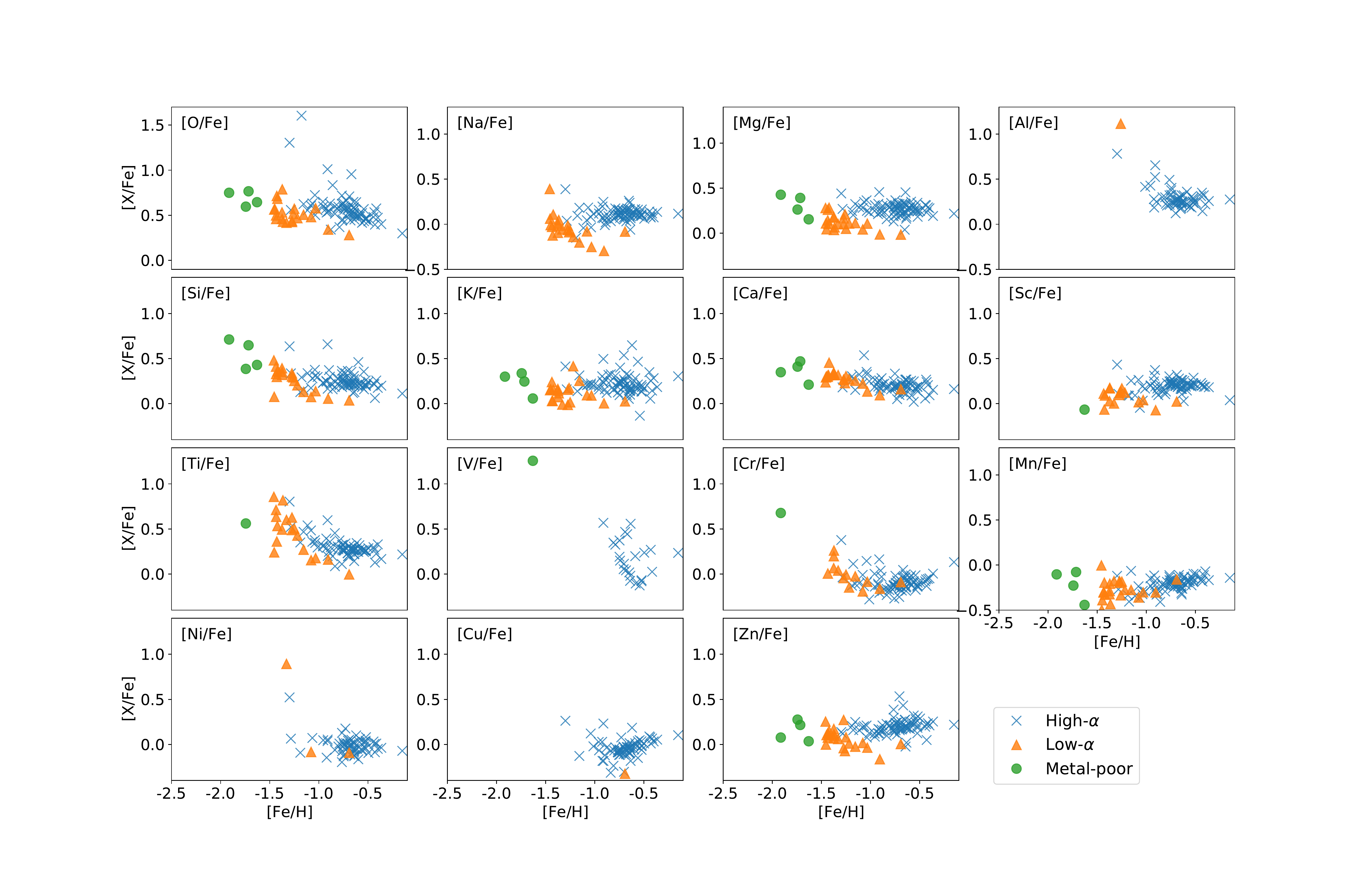}
    \caption{Abundance ratios ([X/Fe]) plotted against 
    [Fe/H] for the three 
    OHS subgroups (``high-$\alpha$", ``low-$\alpha$" and ``metal-poor") defined in 
    Section \ref{sec:chemgroup}. }
    \label{fig:xfe_feh}
\end{figure*}

Figure \ref{fig:xfe_feh} shows distributions of the OHS in [X/Fe] versus [Fe/H] diagrams. As can be seen, the OHS are widely distributed in [Fe/H] and [X/Fe]. A subset of the stars clearly exhibit relatively low [Mg/Fe] ratios at [Fe/H]$\sim -1.0$, similar to the previously reported low-$\alpha$ population in the solar neighborhood \citep[e.g.,][]{nissen10, ishigaki12,hawkins15,hayes18}. The large dispersion in [X/Fe] and [Fe/H] among the OHS implies that they have diverse birth environment. We therefore divide the OHS into three subgroups according to the distribution in the [Mg/Fe]-[Fe/H] 
plane, inspired by preceding studies \citep[e.g.,][]{hawkins15,hayes18,mackereth19}.

The first group is defined 
as high-[Mg/Fe] stars with [Fe/H]$>-1.5$ (``high-$\alpha$" subgroup; blue crosses in Figure \ref{fig:xfe_feh}). The high-[Mg/Fe] 
stars with halo-like kinematics are often interpreted as 
the the early disk populations that have gained high velocity dispersion as a result of minor mergers under the hierarchical Galaxy formation process \citep[e.g.,][]{bonaca17,bonaca20,haywood18,dimatteo19,naidu20,belokurov20,helmi20}. As can be seen in Figure \ref{fig:orbit}, 
these stars are characterized by large $L_{z}$ compared to 
other stars, which is in line with this interpretation.

The second group is defined as low-[Mg/Fe] 
stars with [Fe/H]$>-1.5$ (``low-$\alpha$" subgroup; orange triangles). As can be seen in Figure \ref{fig:orbit}, 
the phase-space distribution of 
the low-$\alpha$ subgroup stars overlap with  
debris stars of the merger of a single galaxy, 
known as "Gaia-Enceladus-Sausage" (GES) structure 
\citep[][see Section \ref{sec:MWformation}]{belokurov18,helmi18}.
The trend of 
decreasing [Mg/Fe] with increasing [Fe/H] for this 
subgroup could be attributed to a 
lower star formation rate in the progenitor galaxy, 
which has lead to the delayed SN~Ia enrichment of Fe 
significant relative to Mg from 
CCSNe of massive stars \citep[e.g.,][]{fernandez-alvar18}.

Finally, the OHS includes four stars with [Fe/H]$<-1.5$ (``metal-poor" subgroup; green circles).  In general, these stars exhibit a large dispersion in [X/Fe] compared to the two higher-[Fe/H] subgroups.  
It is beyond the scope of the present work to attribute these stars to the known major identified debris of past accretions \citep[see e.g., ][]{naidu20,aguado21,matsuno21}

\section{Yield models} 
\label{sec:yield_models}

\subsection{Chemical enrichment scenarios}

In this paper, we test four different hypotheses about 
origin of metals in each of the OHS, which are summarized in 
Table \ref{tab:models}. We describe underlying assumptions and motivations for these hypotheses below.

\begin{itemize}
    \item {\it Pop~III core-collapse supernovae (Model A)}: 
    In this scenario, 
    similar to \citet{ishigaki18}, 
    metals are assumed to have predominantly come from 
    a single CCSN of a 
    Pop~III stars.
Given the relatively high 
metallicities of the OHS analyzed in this paper, 
it is an extreme assumption that these stars have formed 
out of gas purely enriched by a Pop~III CCSN. 
In particular, the high-$\alpha$ OHS 
subgroup exhibit 
[Fe/H] as high as $-0.5$, 
at which this scenario 
is unlikely even with a tight age constraint as we 
quantify in Section \ref{sec:simulation}. These 
stars, however, may retain a representative 
enrichment pattern of multiple Pop~III CCSNe occurred within 
a small host halo. For the metal-poor OHS subgroup  
with [Fe/H]$<-1.5$, a stochastic chemical 
enrichment in the early Universe 
coupled with highly inhomogeneous nature of SN metal ejecta 
would not completely rule out the Pop~III CCSNe enrichment 
\citep{ritter12,salvadori15,tarumi20}. 
By taking into account this scenario, we 
examine whether the observed patterns of abundance ratios ([X/Fe]) can
 accept or rule-out the Pop~III CCSNe enrichment. 

 \item {\it Pop~III $+$ normal core-collapse supernovae (Model B)}: 
 In this model, we consider a scenario that 
 the ejecta of a
 Pop~III CCSN are mixed with interstellar medium 
 enriched by normal (non-zero-metal) CCSNe.
 Thus in this scenario, 
 the OHS have formed out of gas enriched by both 
 a Pop~III CCSN and normal CCSNe. 
 When the oldest halo stars formed, 
 chemical evolution models generally predict that 
 CCSNe of normal massive stars dominate chemical enrichment. 
 The yields of normal CCSNe depend on both progenitor masses 
 and metallicities \citep{woosley95,kobayashi06,nomoto13}. 
 Variation in the characteristic masses or 
 the initial mass function (IMF) 
 of CCSN progenitors is not clearly known, and thus, 
 the Salpeter-type IMF of the form $M^{-\alpha_{\rm IMF}}$ with $\alpha_{\rm IMF}=2.35$ \citep{salpeter55}, which is motivated from local observations, is 
 assumed in this model.
 Even at the oldest epoch, 
 the metallicity of the 
normal CCSN progenitors may be different among various Galactic 
 environment with different star formation rates
 \citep[e.g.,][]{kobayashi11}. In this scenario (Model B), we fix the IMF to the Salpeter-type IMF, while 
 the metallicities of the 
 normal CCSN progenitors are chosen
 depending on 
 the observed [Fe/H] 
 of the OHS as detailed in Section \ref{sec:ccsnyields}.

\item {\it Normal core-collapse supernovae (Model C)}: 
In this model, we assume that the elemental 
abundances of the OHS 
are predominantly determined by normal CCSNe averaged over an IMF
with a characteristic metallicity. 
 In this scenario, the OHS are considered to have formed after 
 the stochastic chemical enrichment by Pop~III stars quenched  
 and before the significant metal production by 
 SNe~Ia started. 
 In contrast to the Model B, we treat the 
IMF slope $\alpha$ and $Z_{\rm CC}$ as free parameters within reasonable ranges as detailed in Section \ref{sec:ccsnyields}. 
 
 \item {\it Normal core-collapse $+$ Type Ia supernovae (Model D)}: 
 In this model, we assume that the metals in the 
 OHS came from both the normal CCSNe and 
SNe~Ia. 
 This scenario is motivated by a speculation that  
 SNe~Ia could have contributed to the 
 chemical enrichment in the system even at 
 the oldest epoch when the 
 OHS likely formed. 
 In fact, since progenitor systems of SNe~Ia remain  
 controversial, it is not clear when SNe~Ia
 started to contribute to the chemical enrichment in the early Universe at various environments \citep[e.g.,][]{maoz14}. 
  It has been suggested that there may be 
  a metallicity limit on the occurrence 
  of SNe Ia in single-degenerate systems \citep{kobayashi98,kobayashi09}, but the OHS selected in this paper 
  includes stars with higher metallicities than the limit.
In the following analysis, we 
consider
SNe~Ia yields of both near-$M_{\rm Ch}$ and sub-$M_{\rm Ch}$ white dwarf progenitors from recent calculations
by \citet{leung18,leung20},
as the origins 
of metals in the OHS,  together with the IMF-averaged normal CCSNe yields.
   We describe details of the SNe~Ia yield models in Section \ref{sec:typeIamodel}.
\end{itemize}

For completeness, we have additionally tested 
whether a combination of a Pop~III CCSN and SNe~Ia 
better explains the observed abundances. We have found that 
quality of the fit tends to be lower than Models 
A-D and, in most cases, the maximum possible contribution of 
SNe~Ia 
is lower than 10 \%. We therefore 
concentrate on 
Models A-D in this paper. 

We should note that the Models A-D do not take into account AGB stars 
as a source of elements detected in the OHS. 
Galactic chemical evolution models predict that 
AGB stars are important sources for  
C, N and neutron-process elements and can contribute after $\sim 30$ Myr after the onset of the Galaxy formation \citep{kobayashi11}. 
For the stars analyzed in this paper, only Y and Ba 
abundances were 
reliably measured among the elements likely produced 
by the AGB stars. We therefore restrict our analysis 
to elements from O to Zn, for which contributions from 
AGB stars to the chemical evolution is expected to 
be small \citep{kobayashi20a}. We separately discuss 
implications from observed Y and Ba abundances in Section \ref{sec:othersources}.

In the following subsections, we describe the 
nucleosynthesis yields
used in this paper in detail. 

\begin{center}
\begin{table*}
\begin{threeparttable}
    \begin{tabular}{lcccc}
    \hline
    Nucleosynthesis source &  Model ID & Free parameters   & Fixed parameters & Reference  \\ \hline
     Pop~III CCSNe    &  A & $M$, $E_{51}$, $M_{\rm mix}$, $f_{\rm ej}$, $M_{\rm H}$ &   - & (1),(2)  \\
     Pop~III CCSNe $+$ Normal CCSNe & B&  $M$, $E_{51}$, $M_{\rm mix}$, $f_{\rm ej}$, $M_{\rm H}$, $f_{\rm CC}$& $\alpha_{\rm IMF}$\tnote{a}, $Z_{\rm CC}$\tnote{b}  & (1),(2),(3)  \\
 Normal CCSNe  & C &$\alpha_{\rm IMF}$, $Z_{\rm CC}$ &  & (3) \\
 Normal CCSNe $+$ SNe Ia &  D& $\alpha_{\rm IMF}$, $Z_{\rm CC}$, $f_{\rm Ia}$, $f_{\rm Ch}$\tnote{c} &  & (3),(4),(5)\\ \hline
    \end{tabular}
      \caption{Description of the four hypothesis (Models A-D) 
    about the origin of metals in the OHS. Reference: (1) 
    \citet{tominaga07}, 
    (2)\citet{ishigaki18}, (3)\citet{nomoto13}, (4)\citet{leung18}, (5)\citet{leung20}}
    \label{tab:models}
    \begin{tablenotes}\footnotesize
\item[a] The IMF slope is fixed at $\alpha_{\rm IMF}=2.35$ for the model B and is changed within $-1\le\alpha_{\rm IMF}\le 3$ for the model C. 
\item[b] The metallicity of CCSNe, $Z_{\rm CC}$, is fixed in Model B according to [Fe/H] of each star (see text for the adopted values). For Models C and D,
the value is changed within $0\le Z_{\rm CC}\le Z_{\rm obs}$, where $Z_{\rm obs}$ is the metallicity corresponding to [Fe/H] of each star, calculated assuming $Z_{\odot}= 0.0152$ \citep{caffau11}.
\item[c] We tested $f_{\rm Ch} = 0.0, 0.2, 0.5$ and $1.0$.
\end{tablenotes}
\end{threeparttable}
\end{table*}
\end{center}

\subsection{Pop~III CCSN \label{sec:PopIIImodels}}

We use the same grid of Pop~III supernova yield models as has been 
used to fit the sample of $\sim 200$ EMP stars in 
I18. The yield models include progenitor masses (13, 15, 25, 40, 100$~M_\odot$) and explosion 
energies, $E_{51}=E/10^{51}{\rm [erg]}=0.5$ for the 
13$~M_\odot$ model, 1 for the 13--100$~M_\odot$ models, 
10 for the 25$~M_\odot$ model, 30 for the $40~M_\odot$, 
and 60 for the $100~M_\odot$ model.
The supernova yield models are calculated based on the mixing-fallback model \citep{umeda02,tominaga07} to approximately take into account 
mixing among different layers of elements and their fallback to the central compact remnant, which presumably occurs in an aspherical 
CCSN \citep{tominaga09}. The model employs three 
parameters, $M_{\rm cut}$, $M_{\rm mix}$, and $f_{\rm ej}$, that 
correspond to the inner and the outer boundaries of the mixing 
zone, and the fraction of mass in the mixing zone finally 
ejected to interstellar medium, respectively. 
As has been done in I18, we 
fix $M_{\rm cut}$ at the mass coordinate approximately corresponds 
to the Fe core radius and vary $M_{\rm mix}$ 
and $f_{\rm ej}$ as free parameters. 

For the Model A, we additionally treat 
hydrogen dilution mass as a free parameter, which is determined to 
reproduce the observed values of [Fe/H] within a 
conservative uncertainty of $\pm 0.2$ dex. 
For the Model B, a total Fe yield is calculated 
as the sum of Fe yields of a Pop~III CCSN and an 
IMF- and metallicity-averaged normal CCSNe.

The mixing-fallback model for the calculation of Pop~III CCSN 
yields have been partly motivated by theoretical predictions that 
the massive Pop~III stars have maintained a high rotational velocity at the end of its evolution \citep[e.g.,][]{ekstrom08} that may affect energy and geometry of the supernova explosion.
The yield models used in this work do not self-consistently 
includes nucleosynthesis specific to rotating massive stars. 
In reality, it has been suggested that light elements 
such as N, Na, or Al can be enhanced as a result of 
rotationally induced mixing during the stellar evolution 
\citep[e.g.,][]{meynet10,takahashi14,choplin19}. 
These elements can also be subject to relatively 
large uncertainty 
due to the efficiency of mixing in the stellar interior 
\citep[e.g.,][]{limongi12}.  
Thus, as in I18, a theoretical 
uncertainties of 0.5 dex is adopted for Na and Al, which
reduces the relative weight of these elements in calculating 
the quality of the fit. 
It has been known that the yield models significantly 
under-predict the yields of Sc, Ti and V \citep{nomoto13}, and thus additional 
nucleosynthesis channels are clearly needed to explain observed 
abundances of these elements. We therefore treat the model
predictions of these abundances as lower limits.

\subsection{Normal CCSN \label{sec:ccsnyields}}

We use the grid of yield models from \citet{kobayashi11} and \citet{nomoto13}, which 
contains progenitor masses of $13$, $15$, $20$, $25$, and $40 M_\odot$ with metallicities $Z=0.0, 0.001, 0.01, 0.02$. For each 
metallicity, we take an average of the yields 
over the progenitor masses 
by weighting with an IMF. The IMF of normal CCSN progenitors remain elusive, while available 
observations have found no significant 
deviation from the Salpeter-form IMF, $\psi(m)\propto m^{-\alpha_{\rm IMF}}$, with $\alpha_{\rm IMF}=2.35$ \citep{bastian10} for massive stars.

For the Model B, 
we take a linear 
combination of a Pop~III SN and a normal CCSN
yield model with an additional parameter $f_{\rm CC}$, 
which corresponds to the fraction of yield from normal CCSNe  relative to the total Pop III SN + normal CCSN yield. The normal CCSN yield 
is calculated by averaging over a 
fixed IMF, with a power-low slope of  
$\alpha_{\rm IMF}=2.35$, and interpolated 
at the characteristic metallicity, $Z_{\rm CC}$. 
The value of $Z_{\rm CC}$, is
fixed depending on the 
observed
[Fe/H] of the OHS. Specifically, 
for the metallicities of the CCSN progenitors, we adopt the yields interpolated at $Z_{\rm CC}/Z_\odot=0.011$, $0.067$, and $0.209$, 
for the OHS with [Fe/H]$<-1.2$, $-1.2\le$[Fe/H]$<-0.7$, and [Fe/H]$\ge-0.7$. These values are compatible with inferred 
 metallicities in bulge, thick disk or halo, for the environments in which 
 old Galactic stellar populations formed \citep{freeman02}. 

For the Model C, in contrast to the Model B,  
we change $\alpha_{\rm IMF}$ in the range 
$-1.0\le\alpha\le3.0$ to bracket the values reported by local observations \citep{bastian10}. 
We also treat 
$Z_{\rm CC}$ as a free parameter in the Model C.

\subsection{SN~Ia \label{sec:typeIamodel}}

For the SN~Ia yield models, we use the yields 
of \citet{leung18} 
and \citet{leung20} taking into account updates in \citet{kobayashi20b}, 
\footnote{The yield tables we use in this paper took into account a longer decay time and solar-scaled initial composition (see \cite{kobayashi20b})} with progenitor metallicities of $Z=0.0$, $0.002$, $0.01$, and $0.02$.
\citet{leung18} and \citet{leung20} provide 
grids of SN~Ia yields for near-Chandrasekhar-mass ($M_{\rm Ch}$) and sub-$M_{\rm Ch}$ white dwarf progenitors, respectively. We make use of the model with a progenitor 
masses of 1.37$~M_{\odot}$ and 1.0$~M_{\odot}$, for the near-$M_{\rm Ch}$ and sub-$M_{\rm Ch}$ models, respectively. 
For the 1.37$~M_{\odot}$ white dwarf, we follow \citet{leung18}, who 
used this model as the 
benchmark model of the $M_{\rm Ch}$ white dwarf
because 
it produced the necessary amounts of Mn and Ni
matching with the element trends
in stars in the solar neighbourhood (see also \citealt{kobayashi20a}). 
The 1.0$~M_{\odot}$ white dwarf is chosen because this model produces
$\sim 0.6M_\odot$ $^{56}$Ni in the ejecta, which is the 
typical amount of $^{56}$Ni observed in SN~Ia
\citep{leung20}. 
For the progenitor metallicities of SN~Ia, we assume 
$Z_{\rm Ia}=0.0$ and $0.1~Z_{\odot}$ 
for the OHS with [Fe/H]$<-1$ and $\geq -1$, respectively. 

For the Model D, we introduce the parameter, $f_{\rm Ia}$, 
which corresponds to the number fraction of SNe~Ia
relative to the total number of normal CCSNe and SNe~Ia.

In this study, we assume that the 
number ratio of near-$M_{\rm Ch}$-to-all SN~Ia 
(near-$M_{\rm Ch}$ + sub-$M_{\rm Ch}$) explosions ($f_{\rm Ch}$)
to be 0.0, 0.2, 0.5, or 1.0. The value of $f_{\rm Ch}=0.2$ is suggested to 
reasonably explain the Solar abundance and a local 
cluster of galaxies by \citet{simionescu19}. 
Note that $f_{\rm Ch} \ge 75$ \% is suggested from Galactic chemical evolution models 
\citep{kobayashi20b}.
The 
value of $f_{\rm Ch}=1.0$ corresponds to  
an extreme case where the SNe~Ia that have 
contributed to metals in the OHS all 
originated from the near-$M_{\rm Ch}$ white dwarf
explosions \citep{leung18}.

\subsection{Fitting yield models to observed abundances}

To fit the yield models to observed 
abundances, we simply assume that the observed 
abundances in [X/Fe] are independent 
and that the likelihood of each [X/Fe] are approximated by 
a Gaussian function with a standard 
deviation corresponding to the 
measurement uncertainty. 
Although this is a crude assumption, 
since a true likelihood is unknown and difficult to obtain, 
we restrict our analysis to this assumption.

For the Models A and B, we search for 
the best-fit yield models that minimize $\chi^2$ among the discrete grid of the model parameters as in I18. Under
the above assumption, this is 
equivalent to maximizing the likelihood function.

For the Model C and D, where we consider continuous 
values for the parameter sets, we adopt a 
Markov-Chain Monte-Carlo (MCMC) algorithm to sample posterior probability distributions of the parameters. 
For this purpose, we make use of PyMC3 \citep{salvatier16} adopting  
flat priors for all the fitting parameters in the range described in 
Table \ref{tab:models}.

\section{Results} 
\label{sec:results}
 
In the following subsections, we describe results of 
fitting the parameters of Models A--D to the observed chemical 
abundances in the OHS sample.  The estimated model parameters for all the OHS are given in 
Tables \ref{tab:modelA_results}-\ref{tab:modelD_results}

% Boldface starts

\subsection{Pop~III CCSN yields (Model A) \label{sec:bestfitPopIII}}

\begin{figure*}
    \begin{tabular}{cc}
\begin{minipage}{0.5\hsize}
    \includegraphics[width=8.3cm,trim={0 0 0 18mm},clip]{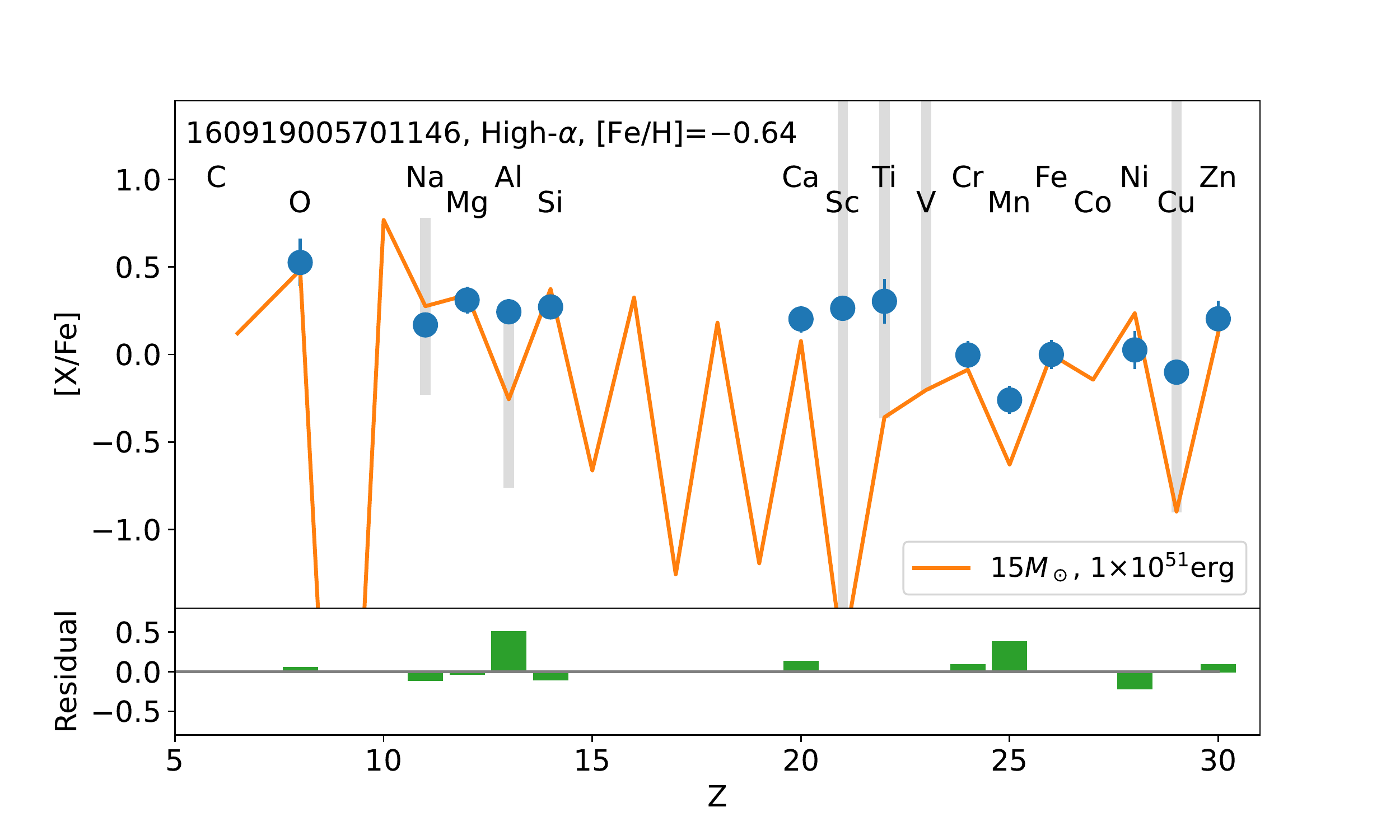}
    \end{minipage} & 
    \begin{minipage}{0.5\hsize}
    \includegraphics[width=8.3cm,trim={0 0 0 18mm},clip]{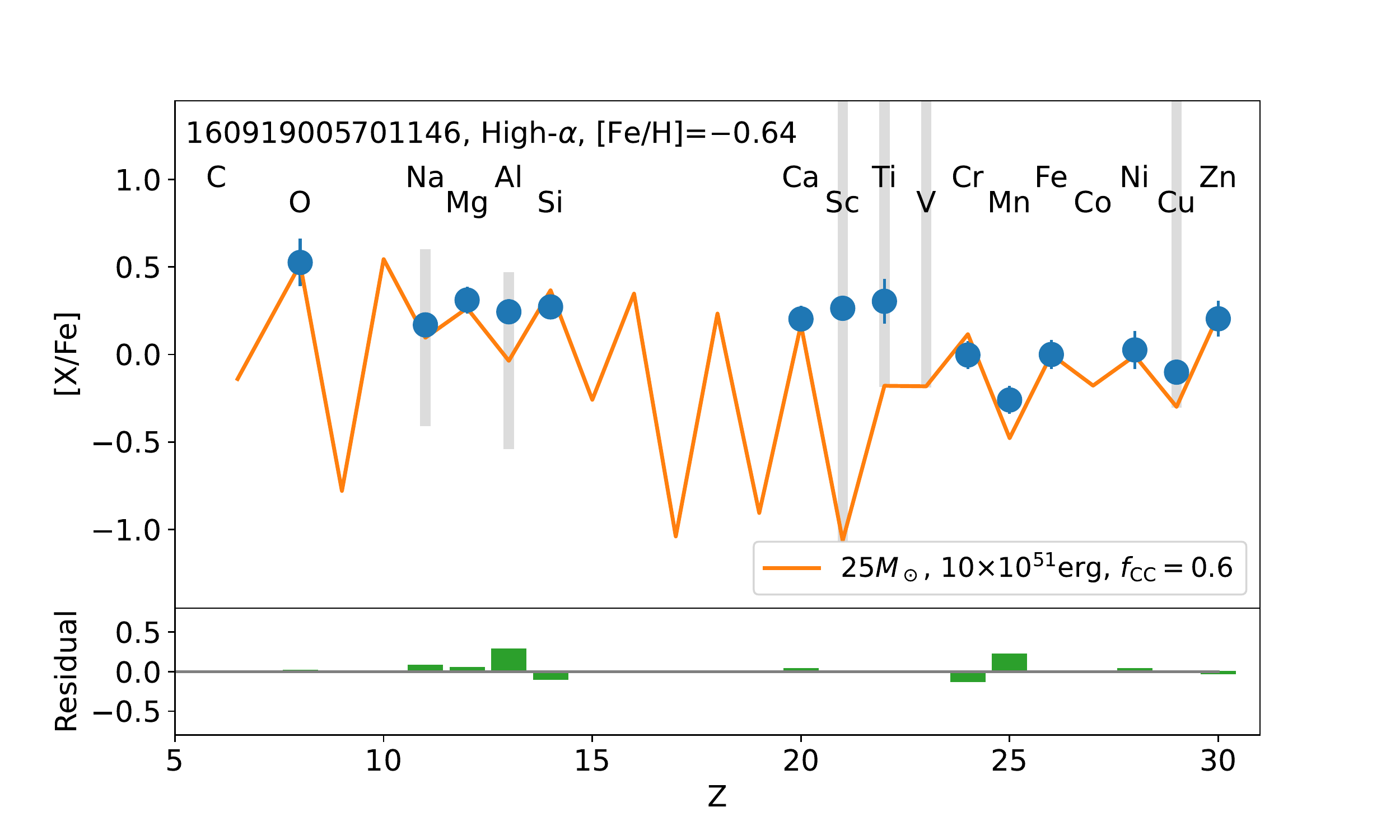}
    \end{minipage} \\
    \begin{minipage}{0.5\hsize}
     \includegraphics[width=8.3cm,trim={0 0 0 18mm},clip]{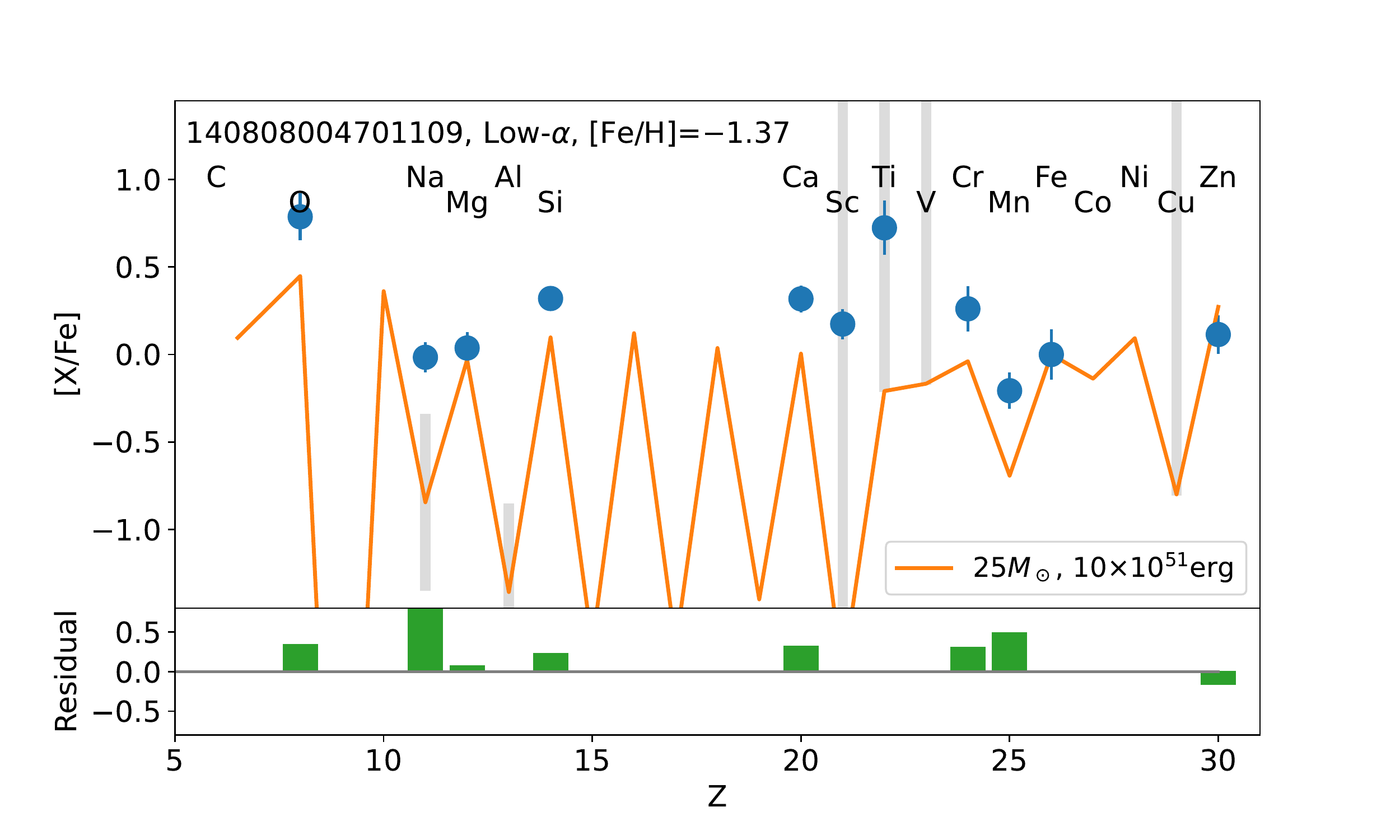}
      \end{minipage} & 
      \begin{minipage}{0.5\hsize}
     \includegraphics[width=8.3cm,trim={0 0 0 18mm},clip]{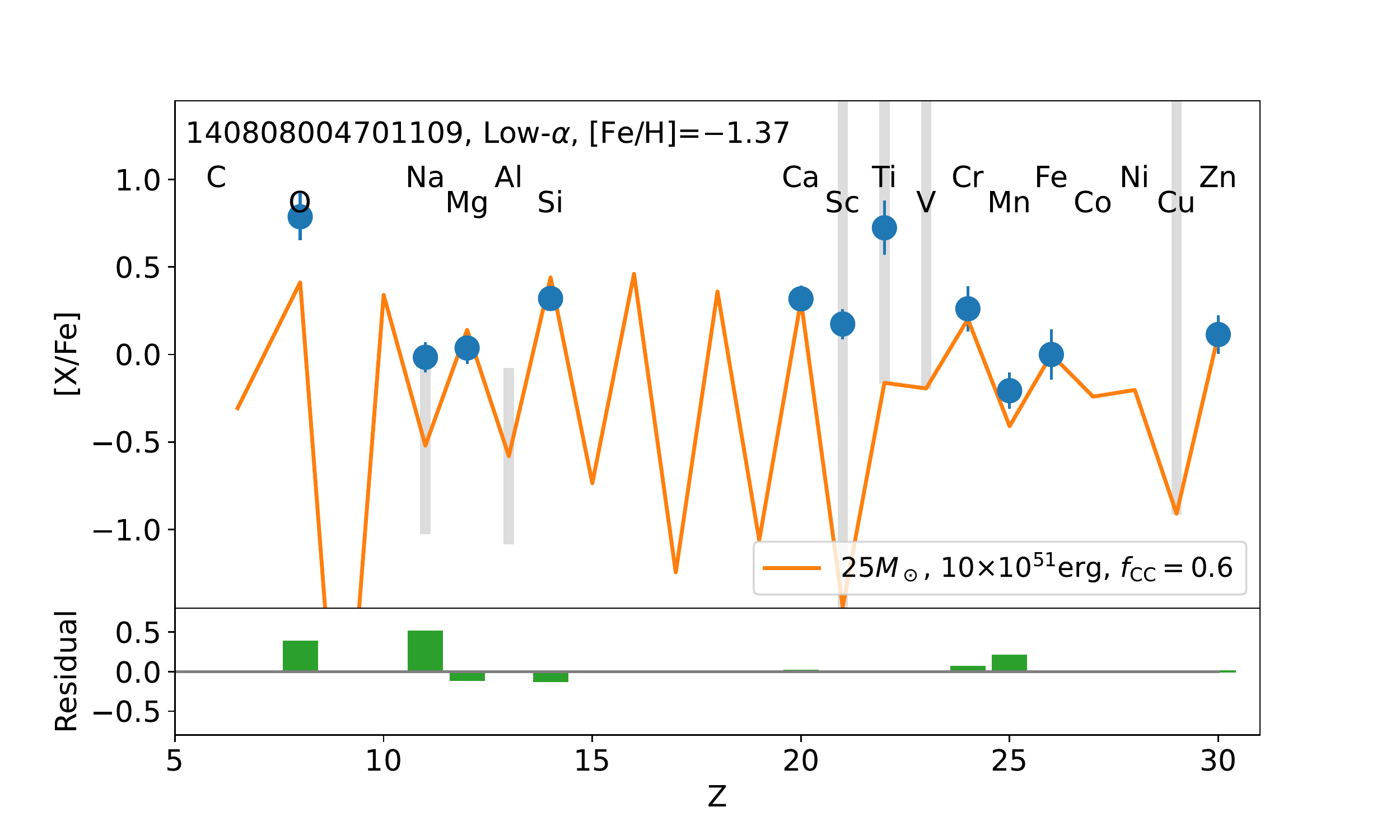}
      \end{minipage}\\
    \begin{minipage}{0.5\hsize}
     \includegraphics[width=8.3cm,trim={0 0 0 18mm},clip]{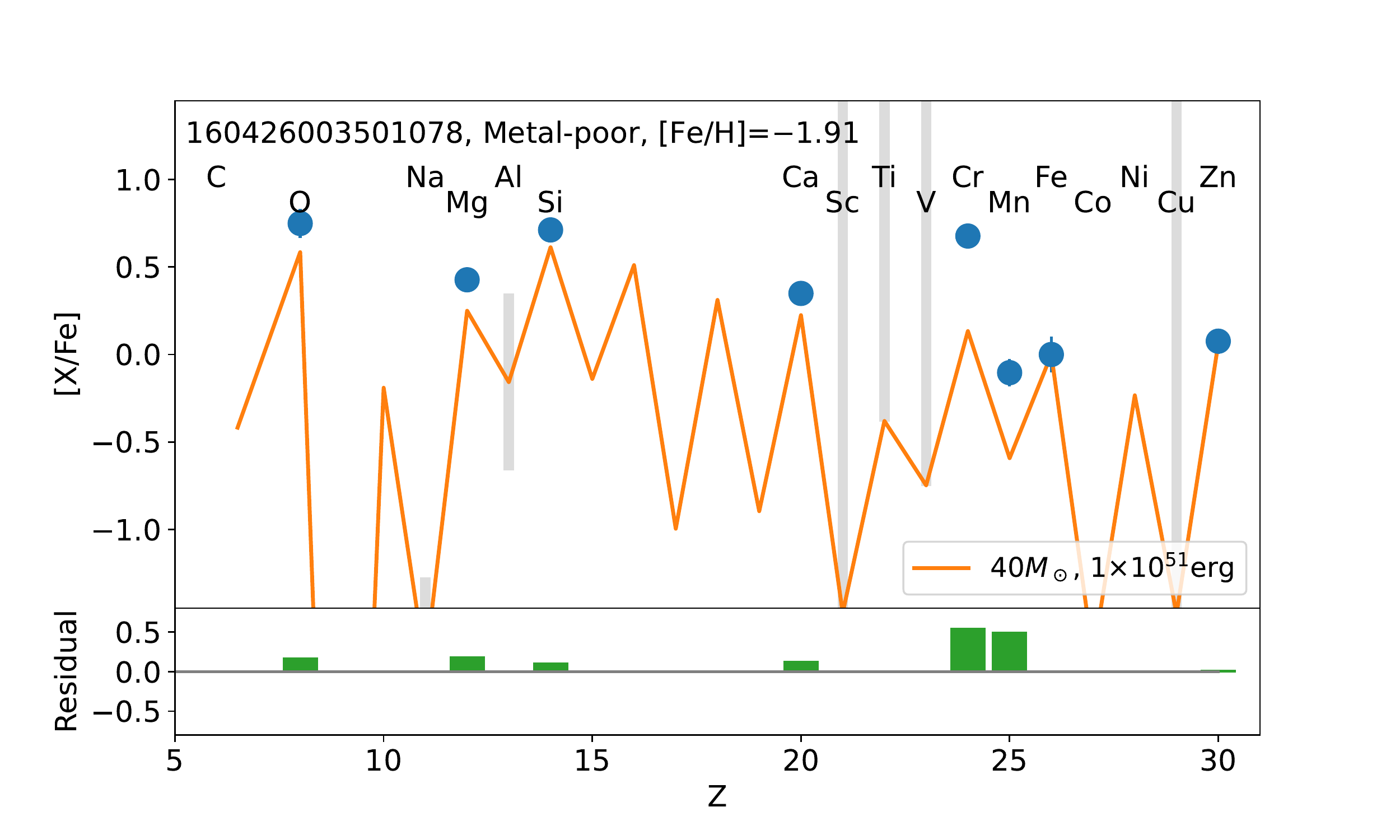}
      \end{minipage} &
      \begin{minipage}{0.5\hsize}
     \includegraphics[width=8.3cm,trim={0 0 0 18mm},clip]{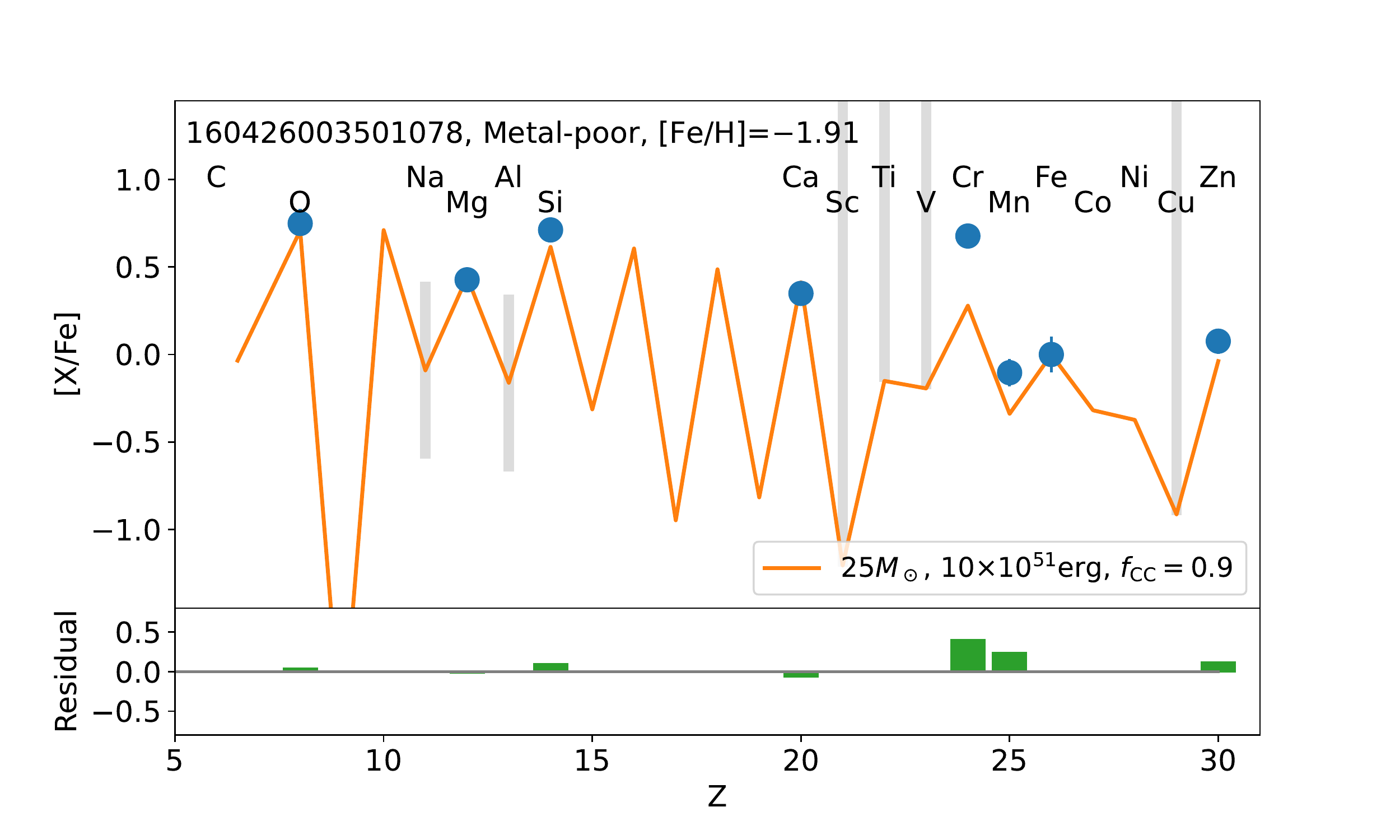}
      \end{minipage}\\
       \end{tabular}
    \caption{The best-fit yield models  (Model A: {\it left} and Model B: {\it right}) compared with observed abundances for the representative stars 
    from the three OHS subgroups. 
From top to bottom, the result for the representative star from the high-$\alpha$, low-$\alpha$ and metal-poor subgroups are shown with GALAH ID, subgroup status and [Fe/H] at the top-left. In each panel, the solid line shows the best-fit yield model, where the corresponding parameter values are shown at the bottom-right. The filled circles correspond to the observed abundances from GALAH DR3. The gray vertical bars mark the elements for which a large theoretical uncertainty is assumed (Na and Al) or the model values are treated as lower limits (Sc, Ti, V, and Cu).  The residuals are shown by solid bars at the bottom of each panel.}
    \label{fig:PopIII_abupattern}
\end{figure*}

The left panels of Figure \ref{fig:PopIII_abupattern} 
show the best-fit Pop III CCSN yield model and 
observed abundances for representative 
stars from the three OHS subgroups. The top 
panel is for one of the high-$\alpha$ OHS, whose 
abundance pattern is characterized by small 
odd-even elemental abundance ratios among 
Na - Ca. Among the yield models considered, 
this pattern is best explained by the Pop~III 
CCSN yield with a progenitor mass of 15$M_{\odot}$. 
The middle panel shows the best-fit model 
for one of the low-$\alpha$ OHS. The low [Mg/Fe] 
of this star is reproduced by the 25$M_{\odot}$ 
Pop~III CCSN yield model. The bottom panel 
shows the best-fit model for one of the metal-poor 
OHS. The high [O/Fe] and [Si/Fe] ratios in this star are reproduced by the 40$M_{\odot}$ 
Pop~III CCSN yield model. For all the 
best-fit models, the abundance ratios of 
odd-Z elements such as Mn are under-produced, 
implying a need for additional metal-enrichment sources for this element. 

\begin{figure*}
    \begin{tabular}{cc}
    \begin{minipage}{0.5\hsize}
    \includegraphics[width=8.5cm]{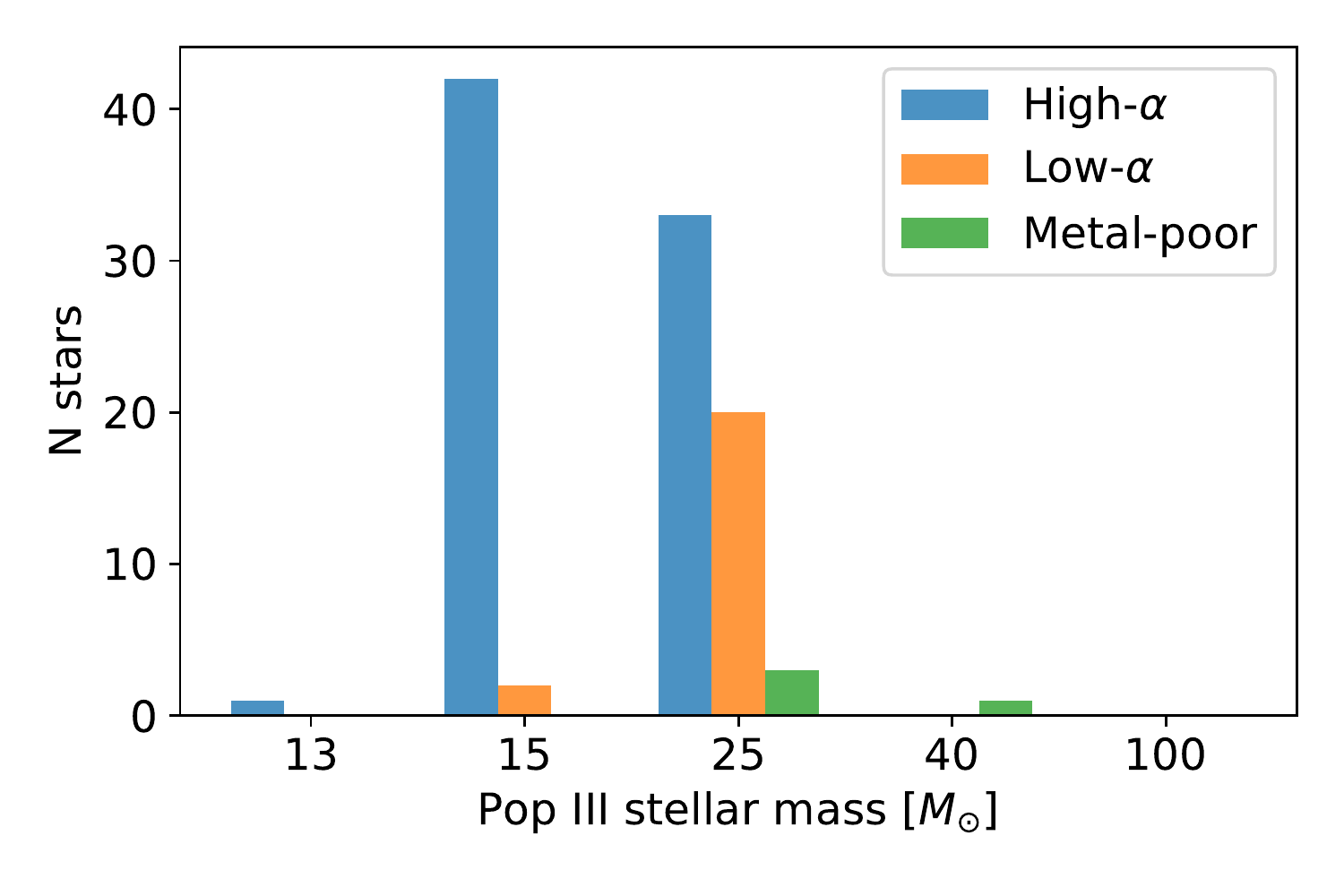}
    \end{minipage} &
    \begin{minipage}{0.5\hsize}
    \includegraphics[width=8.5cm]{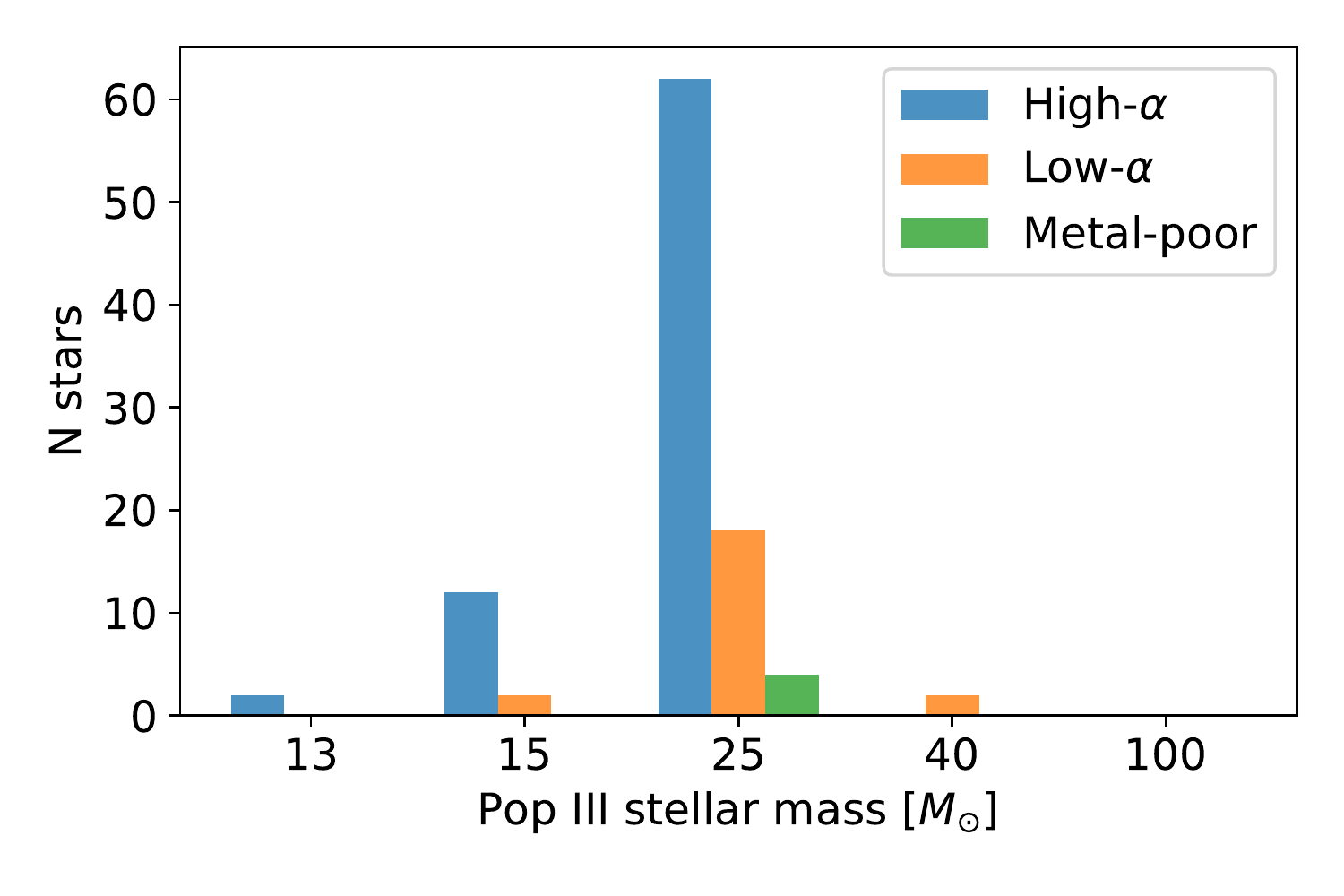}
    \end{minipage}\\
    \end{tabular}
    \caption{The best-fit Pop III stellar masses (13, 15, 25, 40 or 100$M_\odot$) for the Models A({\it left}) and B ({\it right}). The results from the three OHS subgroups are shown by different colors. }
    \label{fig:PopIIImass_hist}
\end{figure*}

The best-fit Pop~III CCSN progenitor 
masses for the three OHS subgroups are 
summarized in the left panel of Figure \ref{fig:PopIIImass_hist}. 
Overall, the observed abundance patterns 
of the OHS are predominantly explained by Pop~III 
CCSN progenitor models of either 15 or 25 $M_{\odot}$.  
A large fraction of the high-$\alpha$ subgroup is best 
fitted by the $15M_{\odot}$ model, while the low-$\alpha$
or metal-poor subgroups are better explained by 
the 25$M_\odot$ model.

The ejected masses of a radio-active 
$^{56}$Ni isotope, which decays to 
$^{56}$Fe, is $\sim 10^{-2}-10^{-1} M_\odot$ for the majority of
the best-fit Pop-III CCSN yield models. 
To explain observed [Fe/H] values of the OHS by a 
single Pop~III yield that fits the abundances, the 
ejected Fe should be diluted with only $10^{2}-10^{3} M_\odot$ of 
hydrogen in most cases (Table \ref{tab:modelA_results}). These values are extremely small compared to simulations of 
metal mixing and analytical models \citep{magg20}. 
We discuss the validity of this scenario in Section \ref{sec:discussion_PopIII}.

\subsection{Pop~III and CCSNe yields (Model B) \label{sec:PopIIICCSNe}}

The right panels of Figure \ref{fig:PopIII_abupattern} 
show the best-fit Pop~III + CCSN yield models (Model B) and 
the observed abundances for the same representative 
stars from the three OHS subgroups as in the left panels.

Compared to the Model A results, quality of 
the fits to the O, Na, Mg, and Al abundance
ratios improve as a result of considering 
the additional contribution from normal CCSNe. 
In each panel, the best-fit fraction of normal CCSNe 
are indicated in the bottom-left corner. The 
metal-poor OHS tends to be better fitted by 
larger fraction of normal CCSNe relative to 
Pop~III SNe. This is partly resulted 
from the low metallicity of the normal CCSNe 
assumed for the metal-poor OHS with [Fe/H]$<-1.5$.

The results of the best-fit Pop III stellar masses are shown in the right panel of Figure \ref{fig:PopIIImass_hist}. Taking into account the contamination 
from normal CCSNe, the majority of the OHS are best fitted by 
the 25$M_{\odot}$ Pop~III CCSN yield models that produce $\sim 10^{-1} M_{\odot}$ 
of $^{56}$Fe (Table \ref{tab:modelB_results}).  Similar to Model A, however, the Fe yields from Pop III CCSN that best explain observed 
abundances violate the constraint on the amount of diluting hydrogen gas to be 
compatible with the observed [Fe/H] \citep{magg20}. 
We discuss the validity of the Models A and B in Section \ref{sec:discussion_PopIII}.

\subsection{Normal CCSN yields (Model C) \label{sec:resultsNormalCCSNe}}

\begin{figure}
    \centering
    \includegraphics[width=8.7cm]{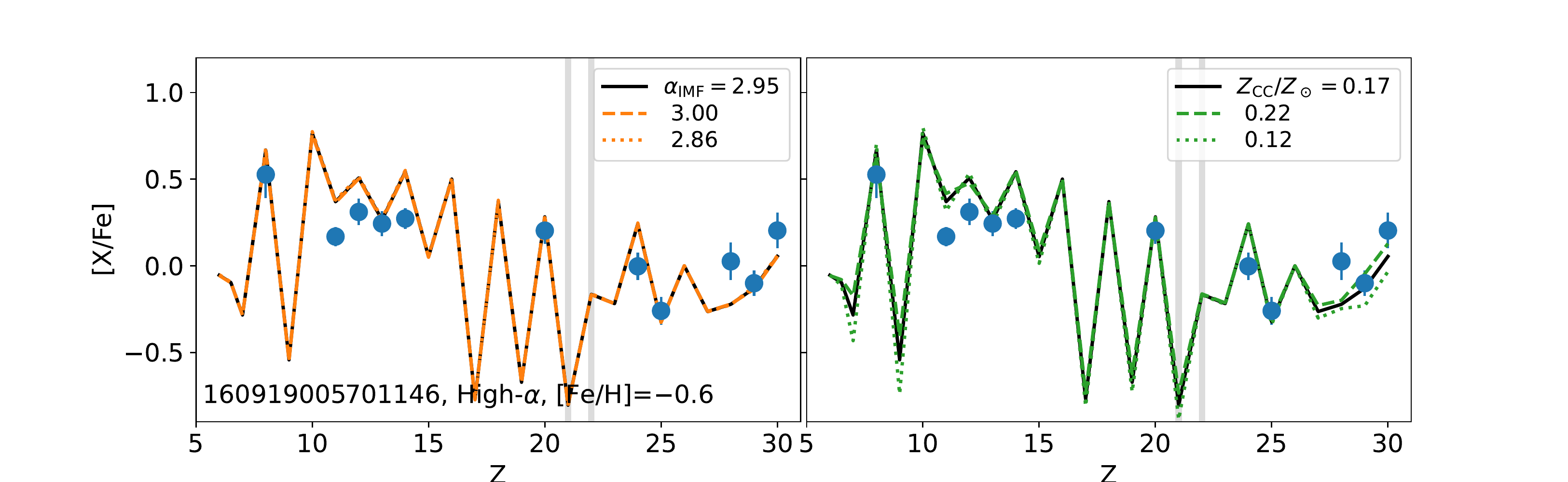}
    \includegraphics[width=8.7cm]{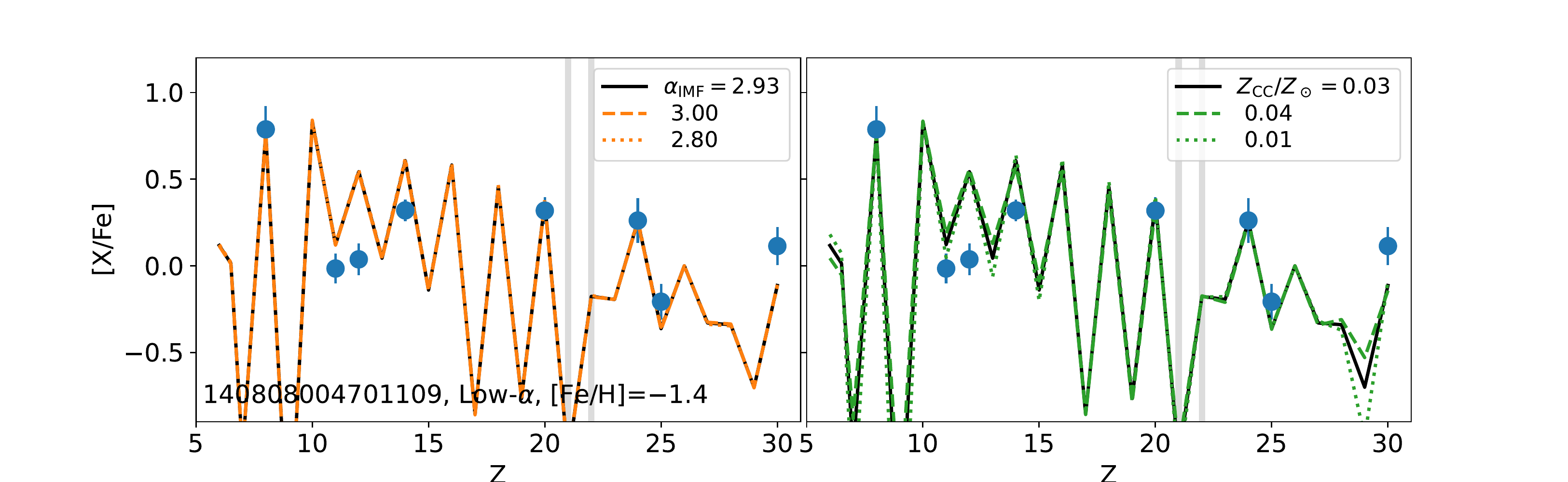}
    \includegraphics[width=8.7cm]{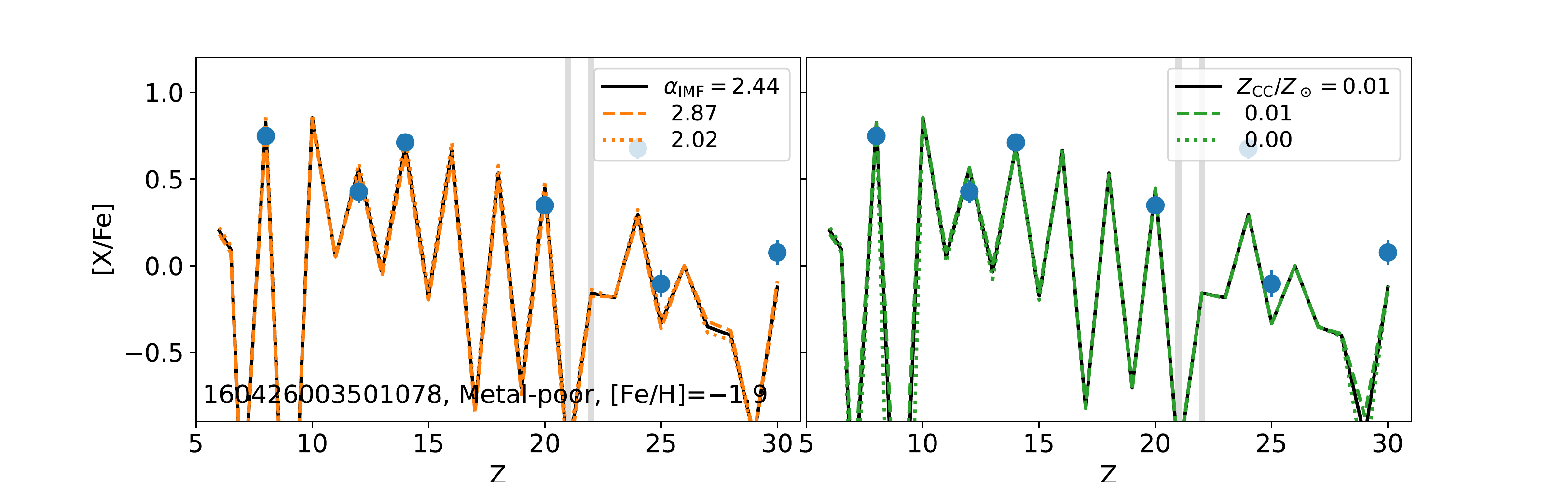}
    \caption{The abundance patterns of IMF-averaged normal CCSN yield models (Model C) compared with observed abundances.  In each panel, the solid line shows the model with the parameters corresponding to the mean of the posterior probability distribution. The circles with error bars are the abundances from the GALAH DR3. 
    The {\it left} and {\it right} panels show the changes in the model abundance pattern by changing $\alpha_{\rm IMF}$ and $Z_{\rm CC}$, respectively. In each panel, the dashed and dotted lines correspond to the lower and higher bounds of the 94 \% highest density interval of the posterior probability distribution. 
   The gray vertical  bands indicate the elements that are not taken into account in the fitting. \label{fig:ccsn}}
\end{figure}

\begin{figure}
 \centering
    \begin{tabular}{cc}
    \begin{minipage}{0.45\hsize}
    \includegraphics[width=4.2cm, trim = 10 10 40 30, clip]{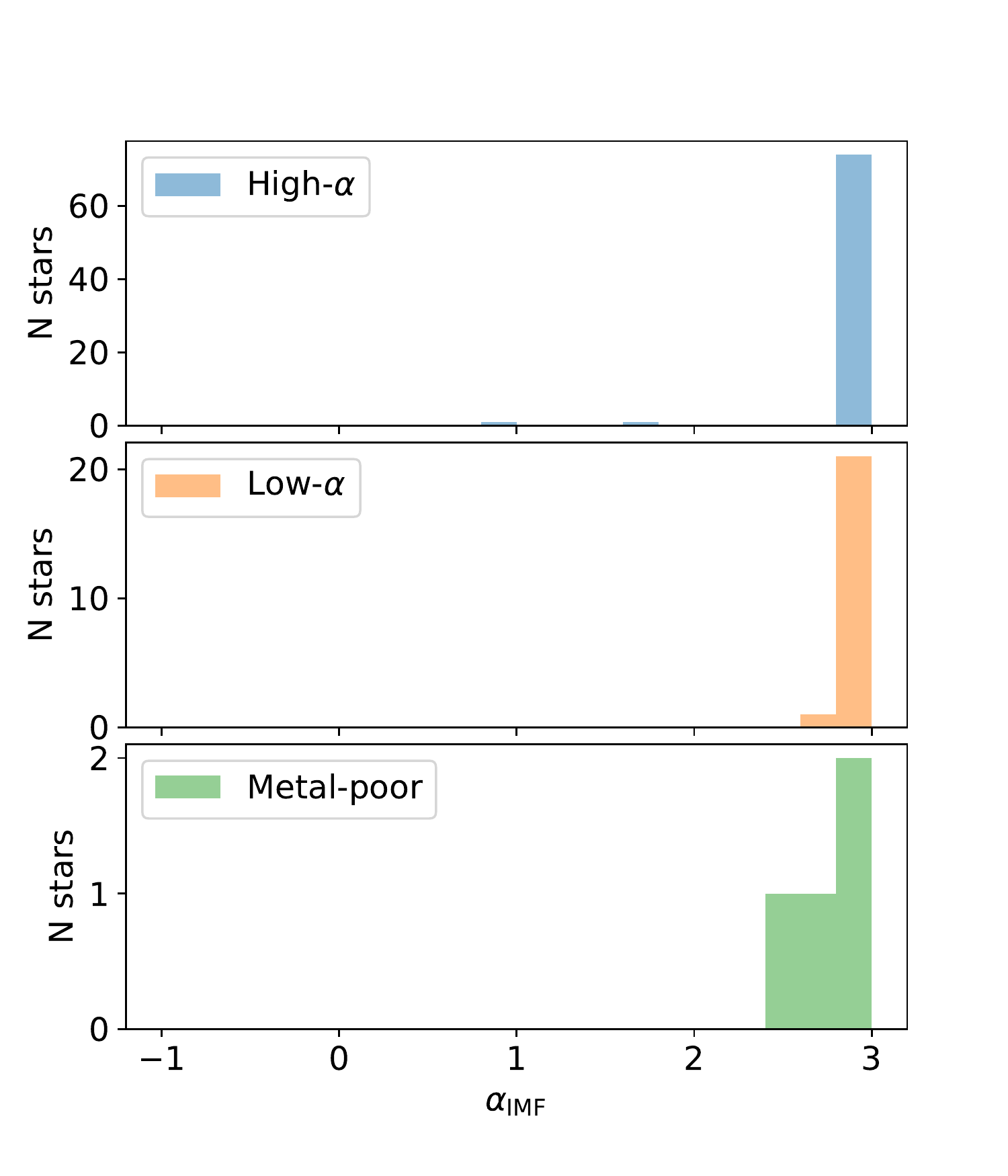}
    \end{minipage} & 
    \begin{minipage}{0.45\hsize}
    \includegraphics[width=4.2cm, trim = 10 10 40 30, clip]{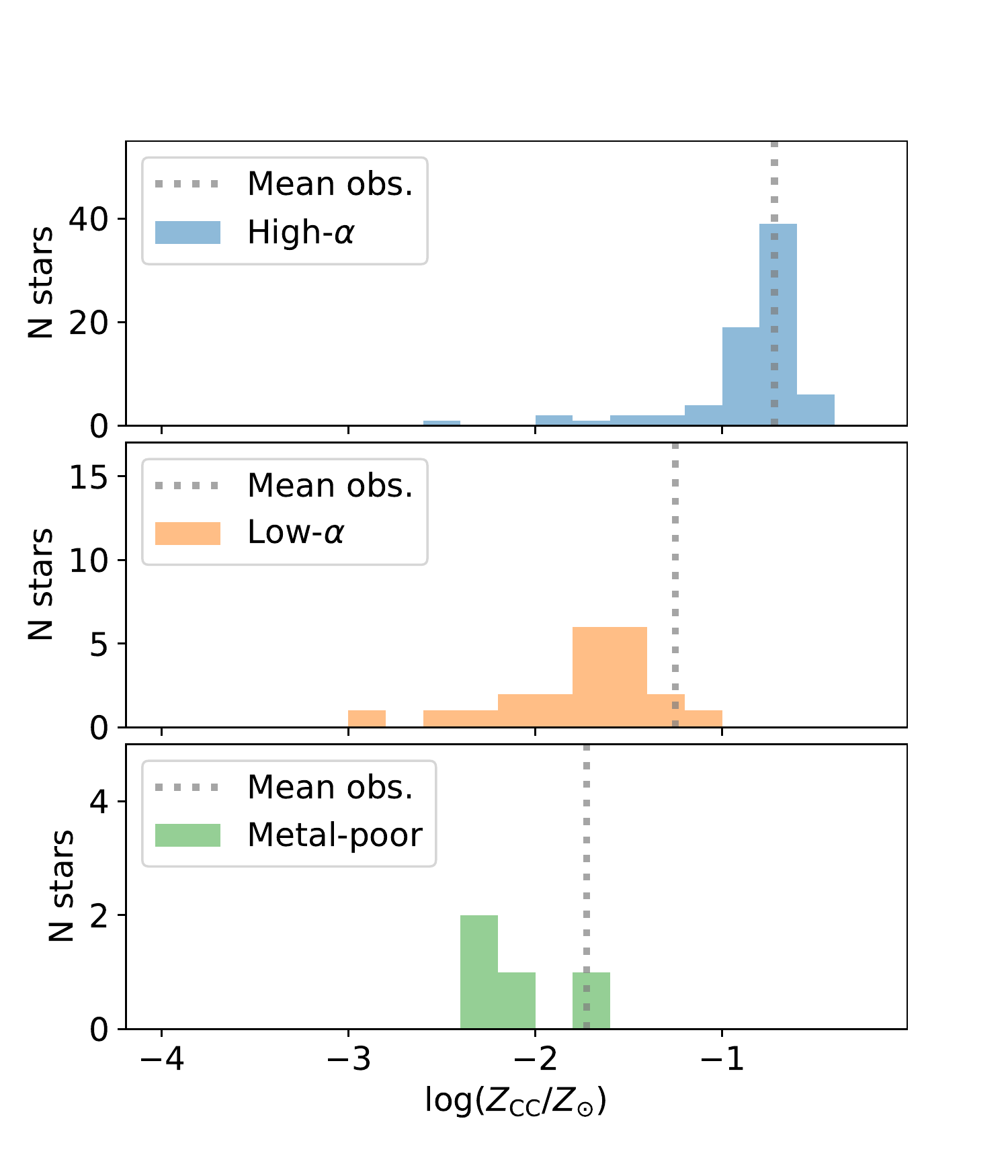}
    \end{minipage} \\
    \end{tabular}
    \caption{Mean values of the posterior probability distributions for the model parameters, $\alpha_{\rm IMF}$ ({\it left}) and $Z_{\rm CC}$ ({\it right}) obtained by the MCMC sampling. 
    The histograms for the 
    three subgroups are shown from top to bottom. In the {\it right} panels, the vertical lines correspond to mean 
    metallicities of the observed stars in each subgroup.}
    \label{fig:hist_CCSN}
\end{figure}

Figure \ref{fig:ccsn} shows results of fitting normal CCSN yields with the IMF slope 
($\alpha_{\rm IMF}$) and the characteristic metallicity of CCSN ($Z_{\rm CC}$) for the 
representative stars from three OHS subgroups.  
The resulting parameter estimates for all the OHS are  
summarized in Figure \ref{fig:hist_CCSN} and in Table \ref{tab:modelC_results}.

To illustrate how the predicted abundances 
change with $\alpha_{\rm IMF}$, 
the left panels of Figure \ref{fig:ccsn} show 
the yield models 
corresponding to the minimum and 
the maximum values that bracket
the 94\% highest density 
interval (HDI) of the posterior 
distribution of $\alpha_{\rm IMF}$, where the HDI provides 
a measure of uncertainty in the parameter estimate.
The mean $\alpha_{\rm IMF}$ value from 
 the posterior distribution 
 is very close to the maximum allowed 
 value for $\alpha_{\rm IMF}$ we have taken into account. 
 The models with smaller $\alpha_{\rm IMF}$ predict 
[X/Fe] of elements from O-Si significantly higher than  
 the observed abundances. 
The larger $\alpha_{\rm IMF}$ corresponds to the 
larger contribution from less massive stars. 
The results thus suggest that, in the context of Model C, the OHS 
are better explained by a larger contribution from 
lower-mass CCSN progenitor stars than expected from the 
Salpeter IMF with the slope of $-2.35$.

Similar to the left panels, 
the right panels of Figure \ref{fig:ccsn} 
illustrate the change in the 
model abundances with $Z_{\rm CC}$. 
The dashed and dotted lines 
indicate the models corresponding to the lower 
and higher bounds of the 94\% HDI 
of the posterior distribution of $Z_{\rm CC}$. 
The change in predicted patterns among different 
$Z_{\rm CC}$ is small for the measured elemental 
abundance ratios. 
As can be seen in Figure \ref{fig:hist_CCSN}, for 
most of the OHS, the mean values of 
$Z_{\rm CC}$ are close to 
the observed metallicity of the OHS 
(dotted lines).

\subsection{Normal CCSN and SN~Ia yields (Model D)\label{sec:resultsCCSNeSNIa}}

\begin{figure*}
    \centering
     \includegraphics[width=16cm]{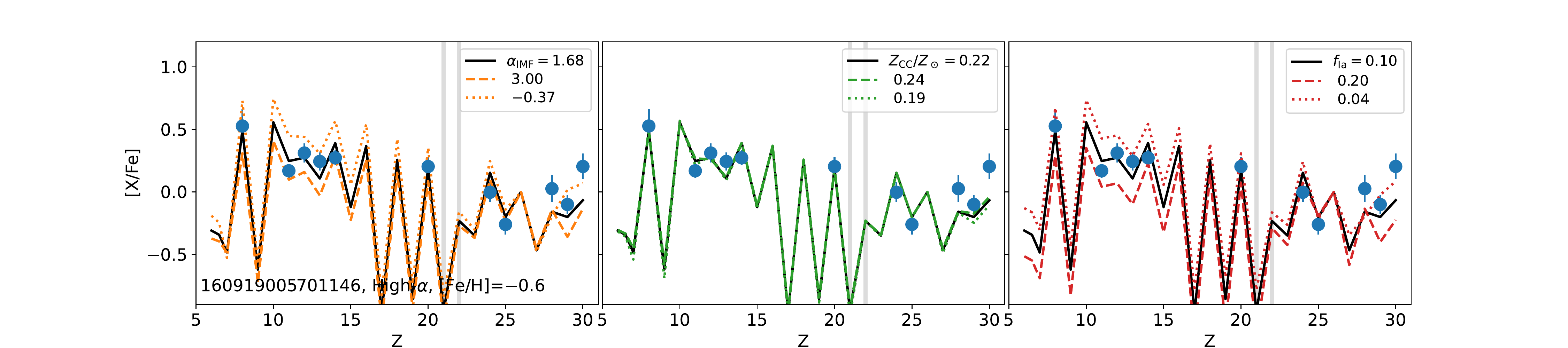}
    \includegraphics[width=16cm]{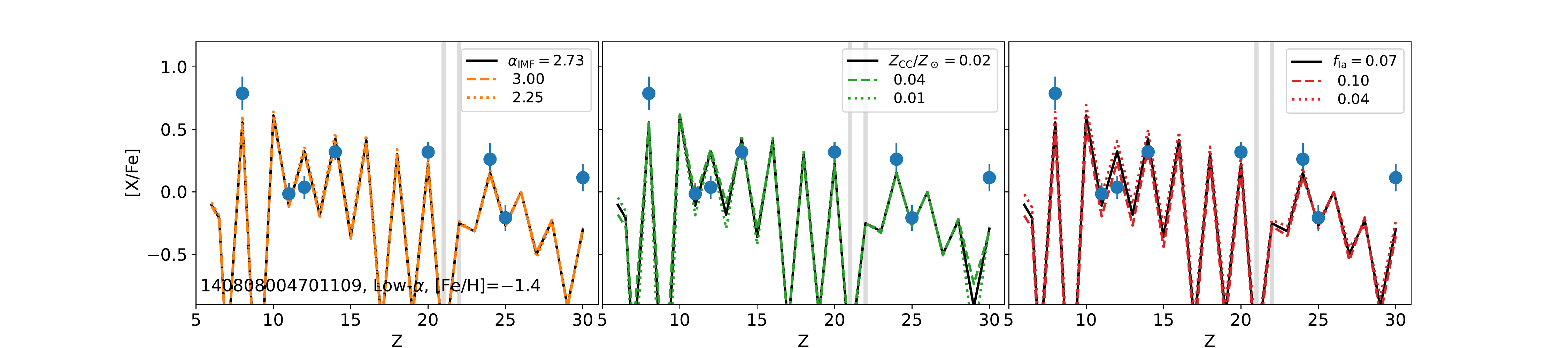}
    \includegraphics[width=16cm]{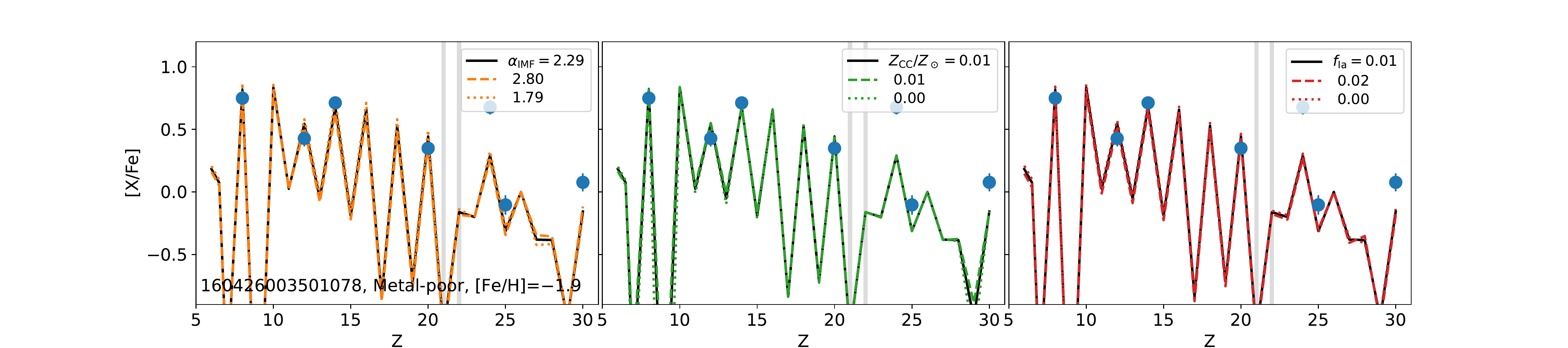}
    \caption{The abundance ratios ([X/Fe]) of the best-fit yield models that assume contributions from both the IMF-averaged normal 
    CCSNe and SNe Ia (Model D). The stars are sorted in order of their metallicity (decreasing from top to bottom). The case for $f_{\rm Ch}=0.5$ is shown. In each panel, the solid line shows the best-fit models and the circles with error bars are the abundances from the GALAH DR2. The models with changing the model parameters, $\alpha$, $Z_{\rm CC}$ and $f_{\rm Ia}$, are shown in the {\it left}, {\it middle} and {\it right} columns, respectively. The gray vertical  bands indicate the elements that are not taken into account in the fitting. \label{fig:ccsn_Ia}}
\end{figure*}

Finally, results of fitting yield models of normal CCSNe
combined with SNe~Ia are shown in Figure \ref{fig:ccsn_Ia} 
for the representative stars of the three OHS subgroups. The result for the case of $f_{\rm Ch} = 0.5$ (equal contributions from 
near-$M_{\rm Ch}$ and sub-$M_{\rm Ch}$ SN~Ia 
progenitors) is shown. The results for all the OHS are given in Table \ref{tab:modelD_results}.  

 Figure \ref{fig:ccsn_Ia} shows abundance patterns corresponding to the mean values of 
the posterior probability distribution of the model parameters. To illustrate 
the parameter dependence of the model prediction, 
different columns show the models with different values of the IMF slope of the CCSN progenitors 
($\alpha_{\rm IMF}$; left), characteristic 
metallicity of the CCSNe ($Z_{\rm CC}$; middle) and the SN~Ia fraction ($f_{\rm Ia}$; right). Compared to the CCSN-only model (Model C, Figure \ref{fig:ccsn}), 
the observed abundance ratios are much better reproduced, 
especially for the high-$\alpha$ and the low-$\alpha$ OHS subgroups.

\begin{figure*}
    \begin{tabular}{ccc}
    \begin{minipage}{0.31\hsize}
    \includegraphics[width=6cm]{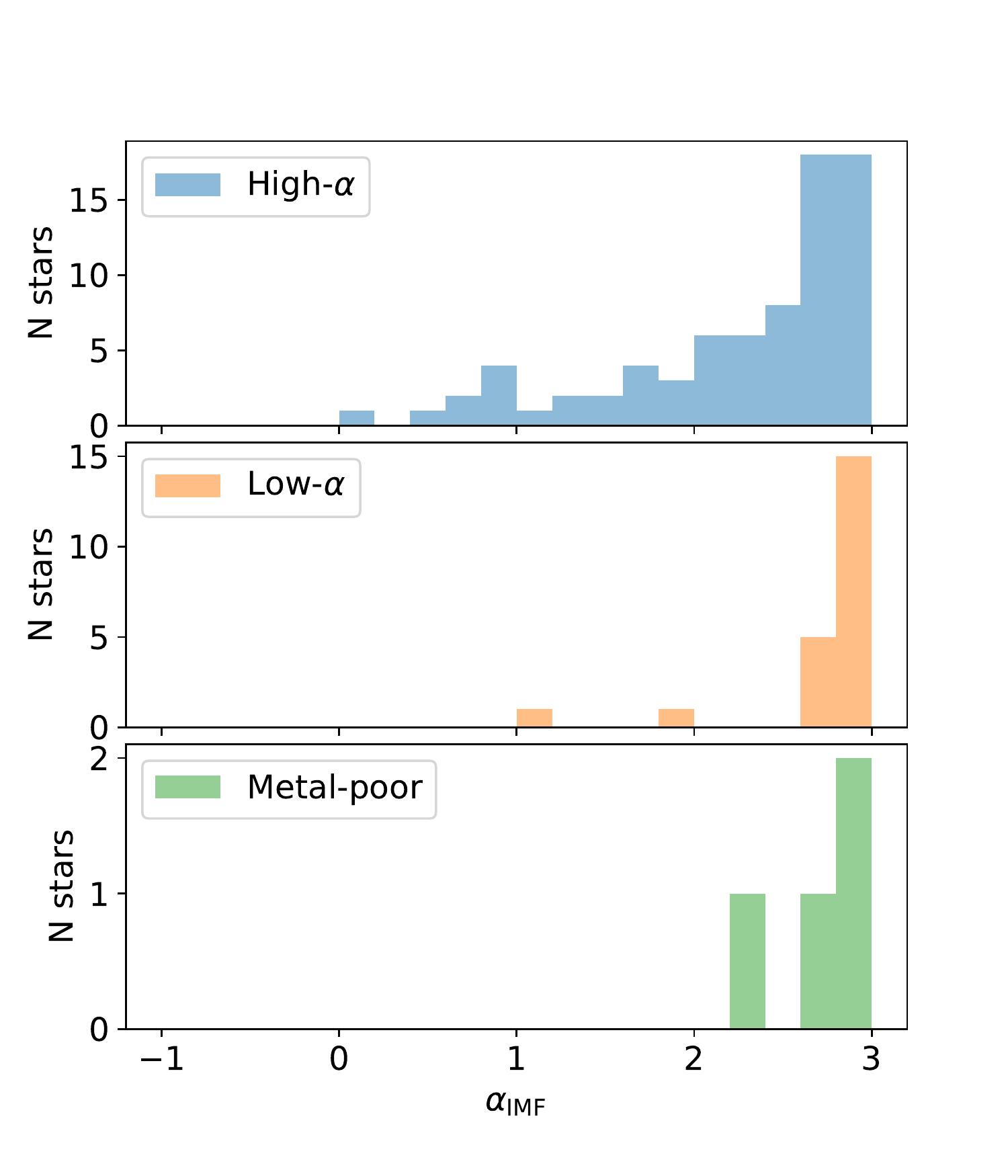}
    \end{minipage} & 
    \begin{minipage}{0.31\hsize}
    \includegraphics[width=6cm]{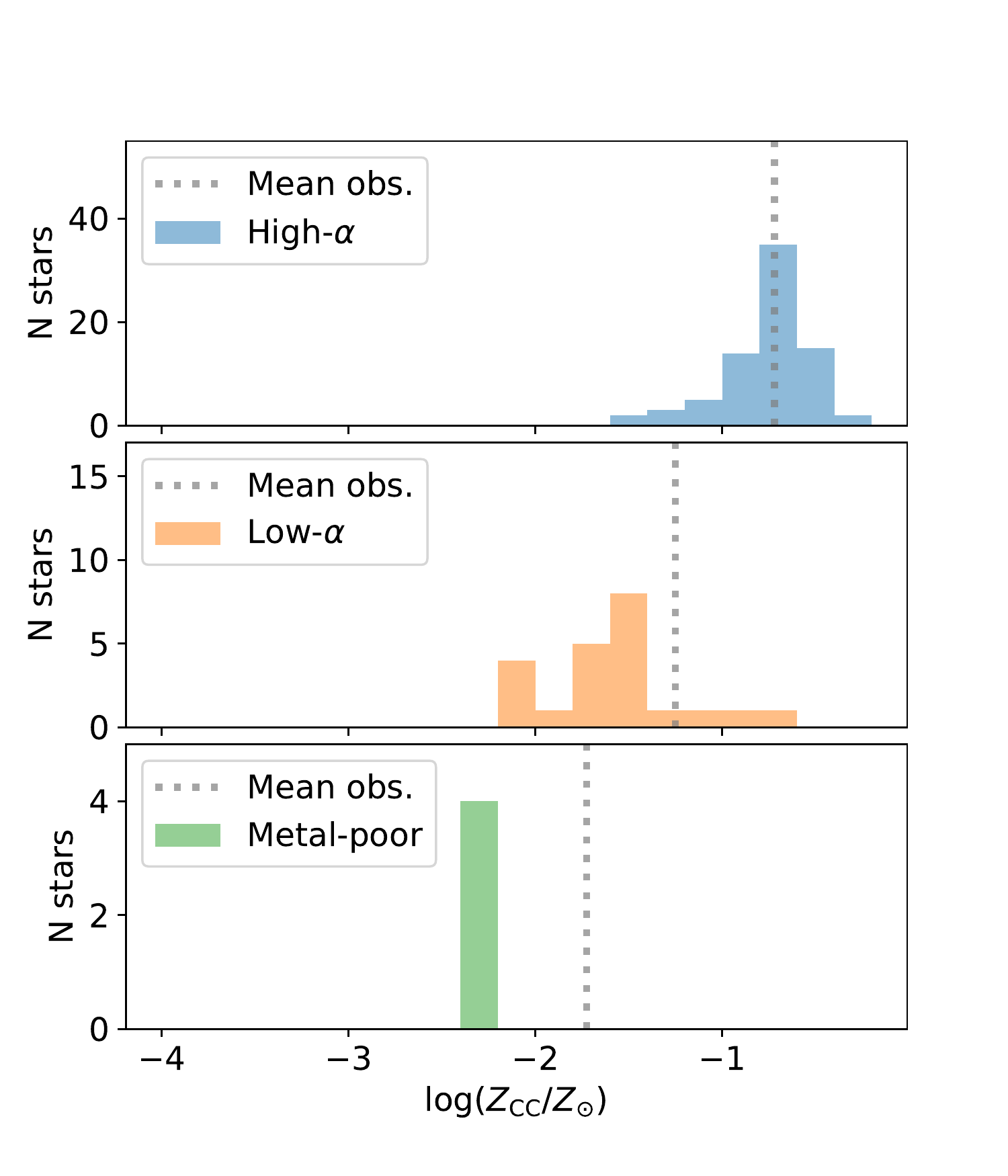}
    \end{minipage}&
    \begin{minipage}{0.31\hsize}
    \includegraphics[width=6cm]{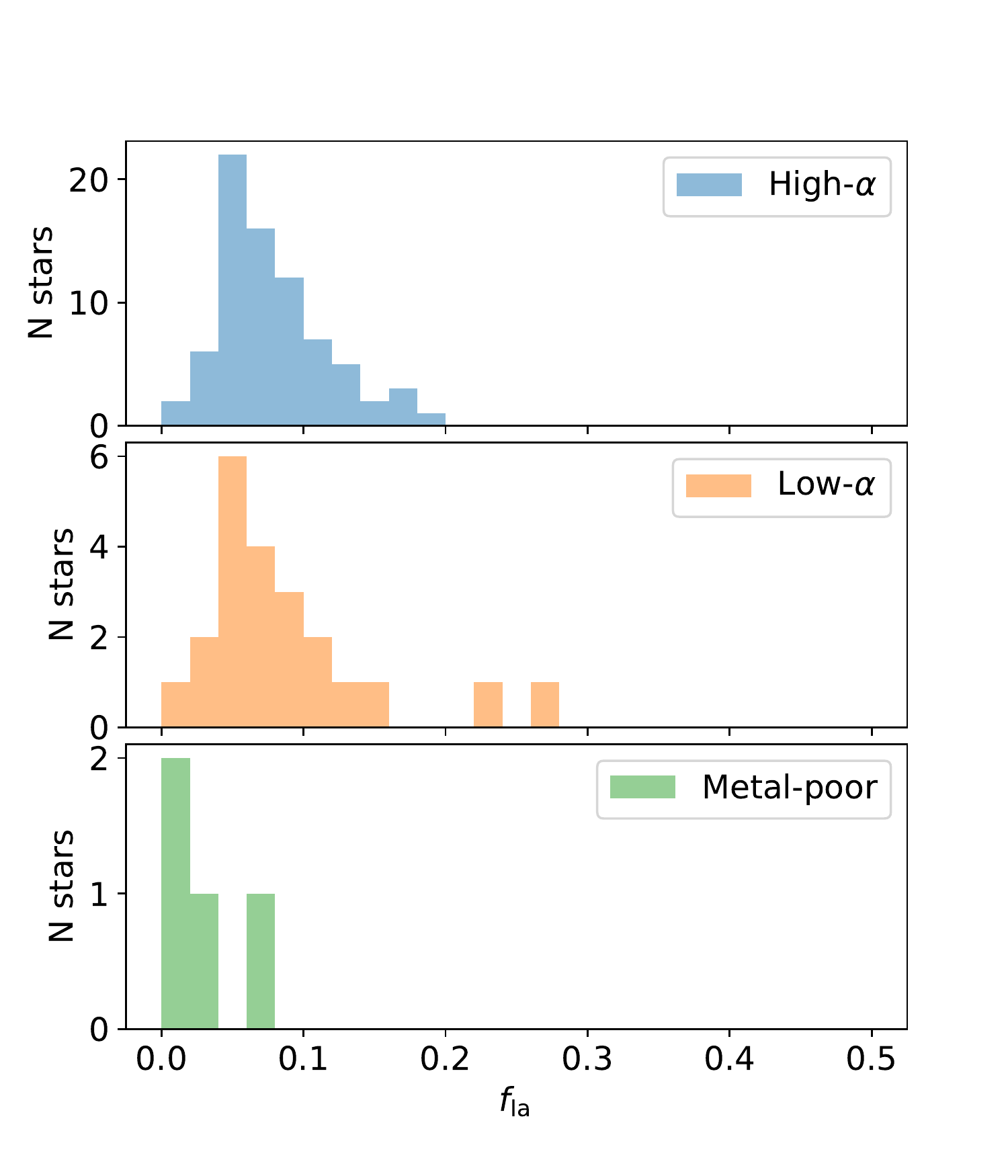}
    \end{minipage}\\
    \end{tabular}
    \caption{
    Mean values of the posterior probability distributions for the model parameters, $\alpha_{\rm IMF}$ ({\it left}), $Z_{\rm CC}$ ({\it middle}) and $f_{\rm Ia}$ ({\it right}) obtained by the MCMC sampling. 
    The histograms for the 
    three subgroups are shown from top to bottom.
    The case for $f_{\rm Ch}=0.5$ is shown. 
    In the {\it middle} panels, the vertical lines correspond to mean 
    metallicities of the observed stars in each subgroup.
    }
    \label{fig:hist_CCSN_Ia}
\end{figure*}

Mean parameter values for $\alpha_{\rm IMF}$, 
$Z_{\rm CC}$ and $f_{\rm Ia}$ from the posterior distributions for each OHS subgroup are 
summarized in Figure \ref{fig:hist_CCSN_Ia} for the 
case of $f_{\rm Ch} = 0.5$. 
Similar to the result of Model C, 
the left panel shows that $\alpha_{\rm IMF}$ 
are mostly distributed 
around the steepest 
possible slope, $\alpha_{\rm IMF}=3.0$, we have taken into account. The 
values of $Z_{\rm CC}$ are in the range 0.03-0.27$Z_{\odot}$, which 
are close to the metallicity of the OHS themselves 
(dotted vertical lines). 
The right panel 
of Figure \ref{fig:hist_CCSN_Ia} shows that 
the values of $f_{\rm Ia}$ 
for the high-$\alpha$ subgroup 
range from 0.00 up to 0.20 (0.08 on average). 
The $f_{\rm Ia}$ values are, on average, 
higher for the low-$\alpha$ 
subgroup (0.01-0.27, 0.09 on average). 
For the metal-poor subgroup, 
the $f_{\rm Ia}$ values are at most $\sim 0.05$,
which are lower than the other two subgroups. 
The SN~Ia fractions depend on 
the different assumptions about $f_{\rm Ch}$. If 
$f_{\rm Ch} = 1.0$ (all SNe~Ia from near-$M_{\rm Ch}$ 
progenitors), instead of $f_{\rm Ch} = 0.5$, is assumed, the SN~Ia fractions 
slightly decrease by a few percent for 
all of the OHS subgroups. 

\begin{figure}
    \centering
    \includegraphics[width=8.5cm]{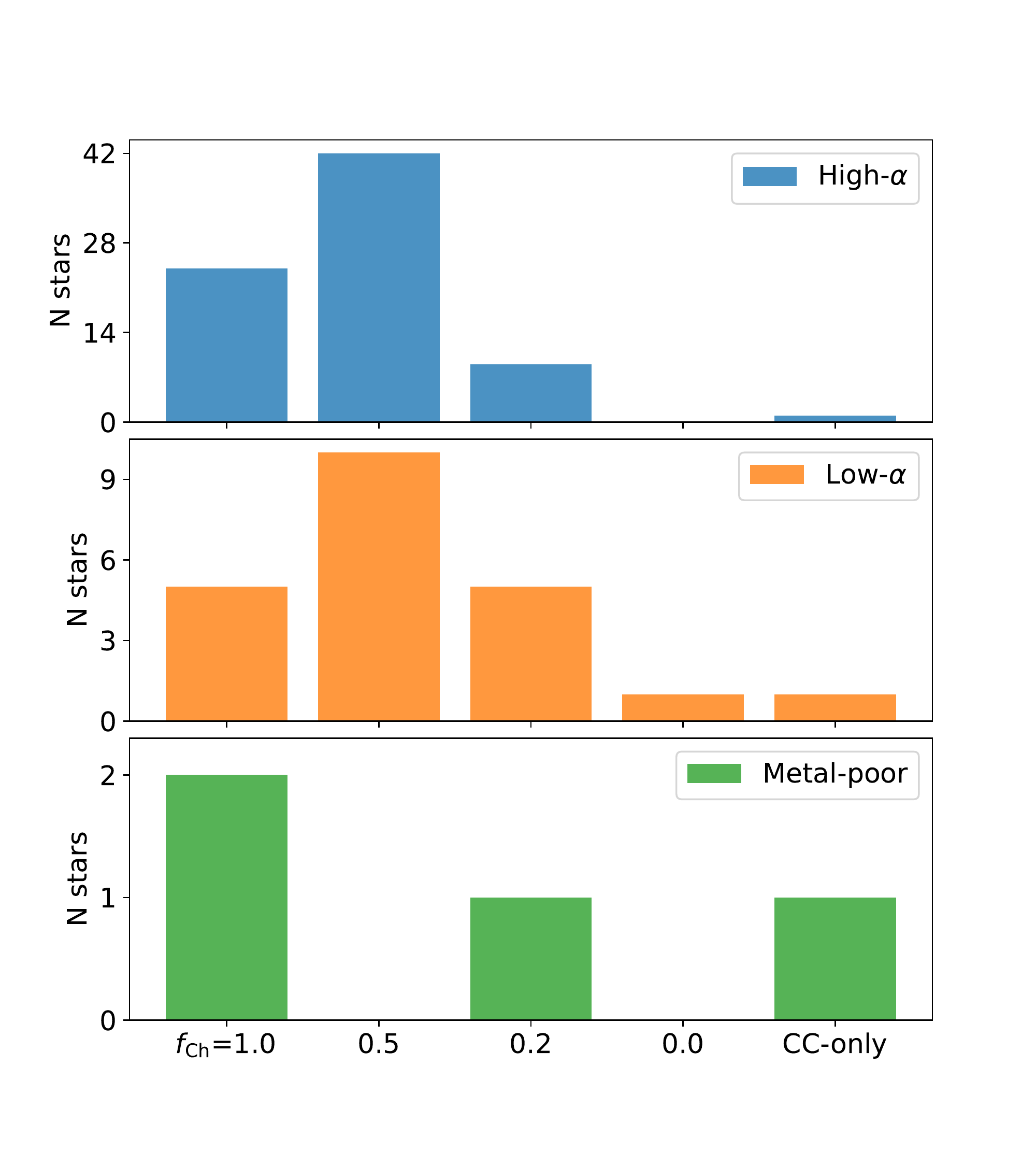}
    \caption{The number of stars that are best explained by Model D with $f_{\rm Ch} = 1.0, 0.5, 0.2$ or $0.0$ (the four bars on the left) or by 
    Model C (the bars on the right). }
    \label{fig:fCh_comparison}
\end{figure}

To compare the models with different values of $f_{\rm Ch}$ and 
with the normal CCSNe only model (Model C),  
for each OHS, we obtain a ranking of these models 
according to the quality of the fit 
by penalizing with the number of parameters. 
Figure \ref{fig:fCh_comparison} summarizes the number 
of stars that are best explained by the models 
with $f_{\rm Ch}=0.0$, $0.2$, $0.5$ or $1.0$  or with 
the normal CCSN only.
 For the 
high-$\alpha$ and low-$\alpha$ OHS,   
we find that the model with $f_{\rm Ch} = 0.5$ 
most frequently best explains the observed abundances.
On the other hand, for the metal-poor OHS, 
the model with $f_{\rm Ch} = 1.0$ 
more frequently best explains the observed  
abundances.

\section{Discussion}
\label{sec:discussion}

Identifying nucleosynthetic sources at various 
galactic environments in the early Universe remains 
a major challenge in studies of the cosmic 
chemical evolution, since it is not feasible to 
directly observe individual stars or SN events 
at high redshifts. 
In our study, we address this question by elemental abundances 
measured by GALAH DR3 \citep{buder21} for 
relatively old Milky Way halo 
stars (``OHS'')) in the solar neighborhood. 
We have considered yield models that represent 
four different hypotheses about the origin of metals 
in the atmosphere of the OHS (Models A-D). For each of the 
Models A-D, we have obtained the model parameters that reproduce 
observed elemental abundances. 

\subsection{Comparison among models \label{sec:discussionComparison}} 

\begin{figure}
    \centering
    \includegraphics[width=8.3cm]{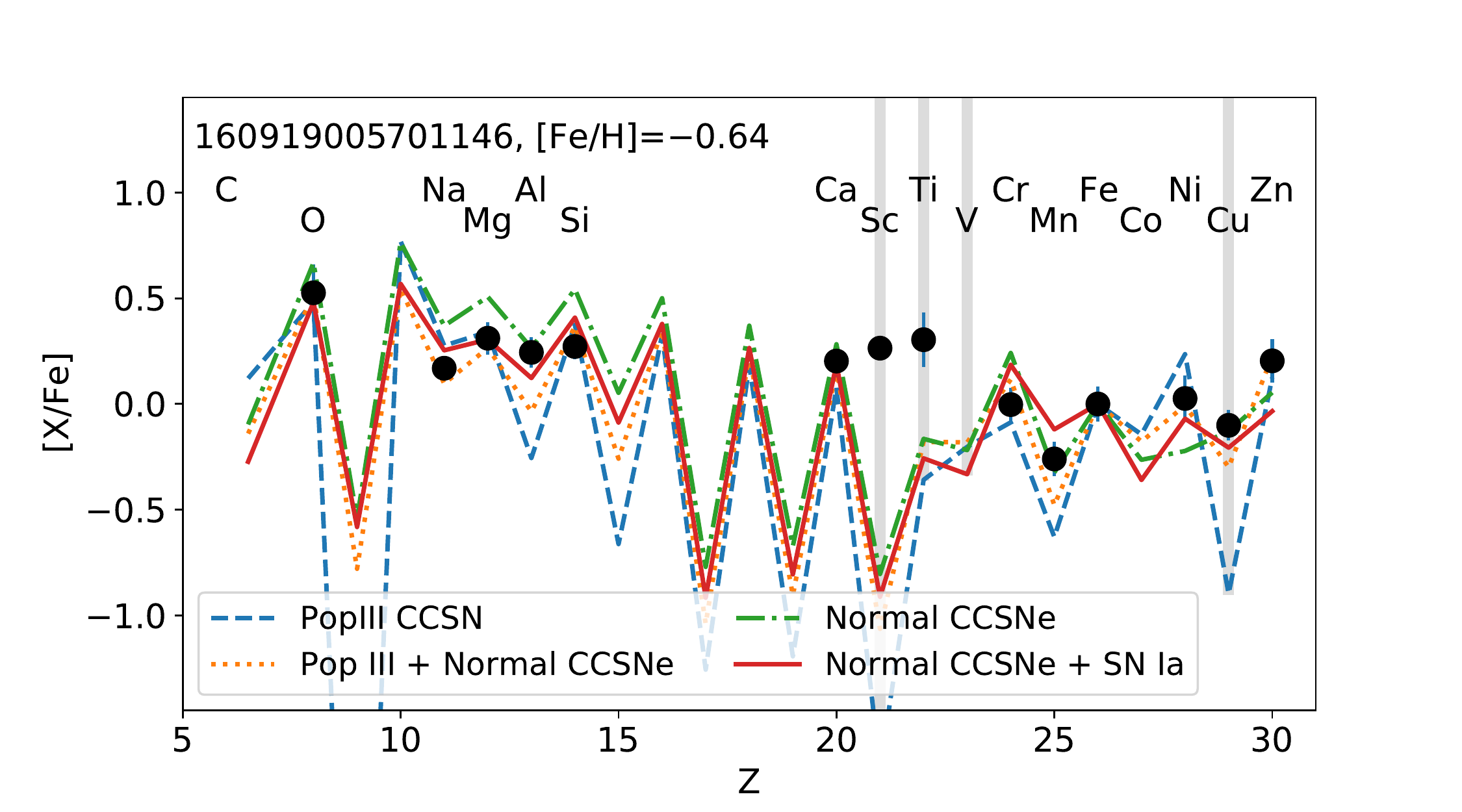}
    \includegraphics[width=8.3cm]{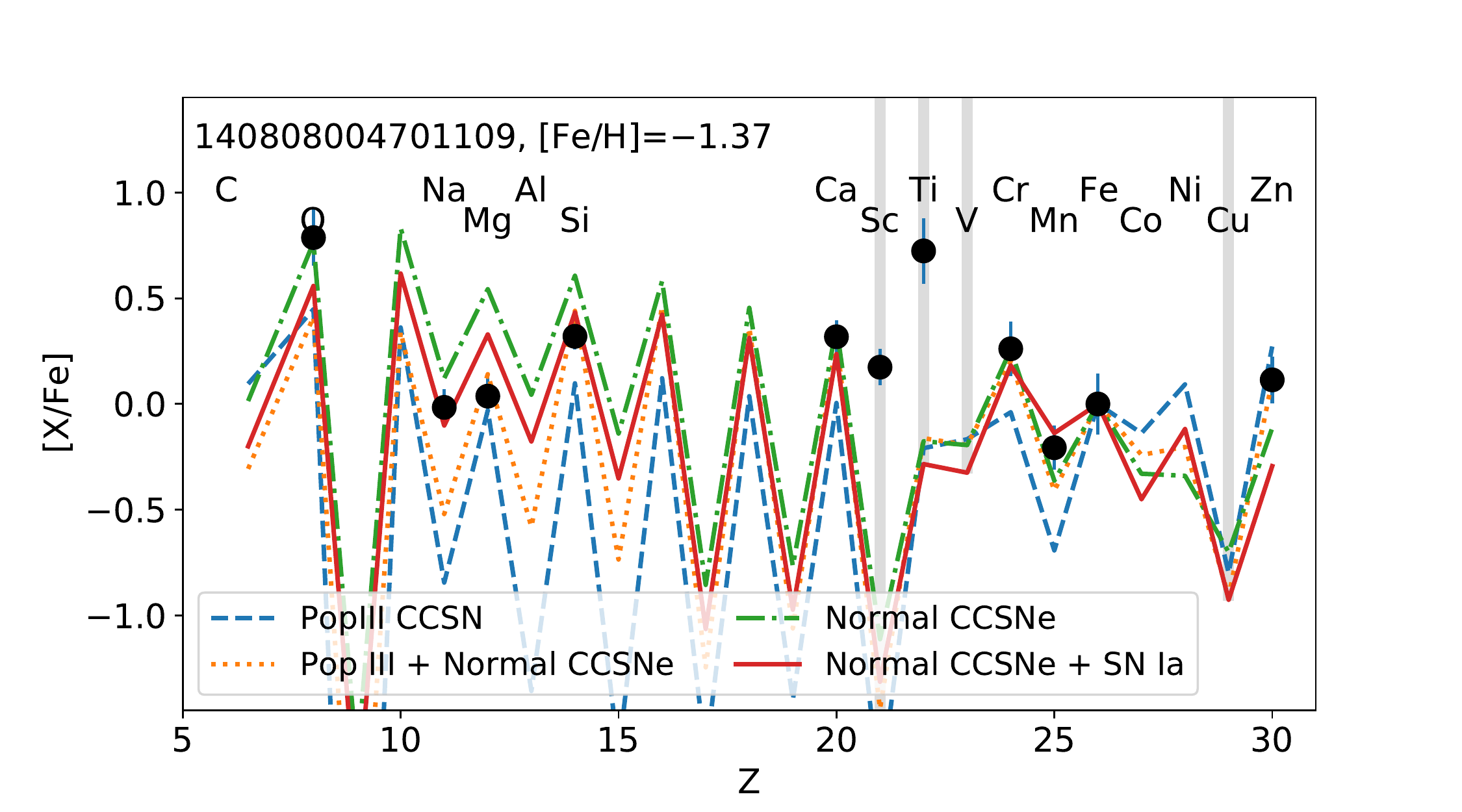}
    \includegraphics[width=8.3cm]{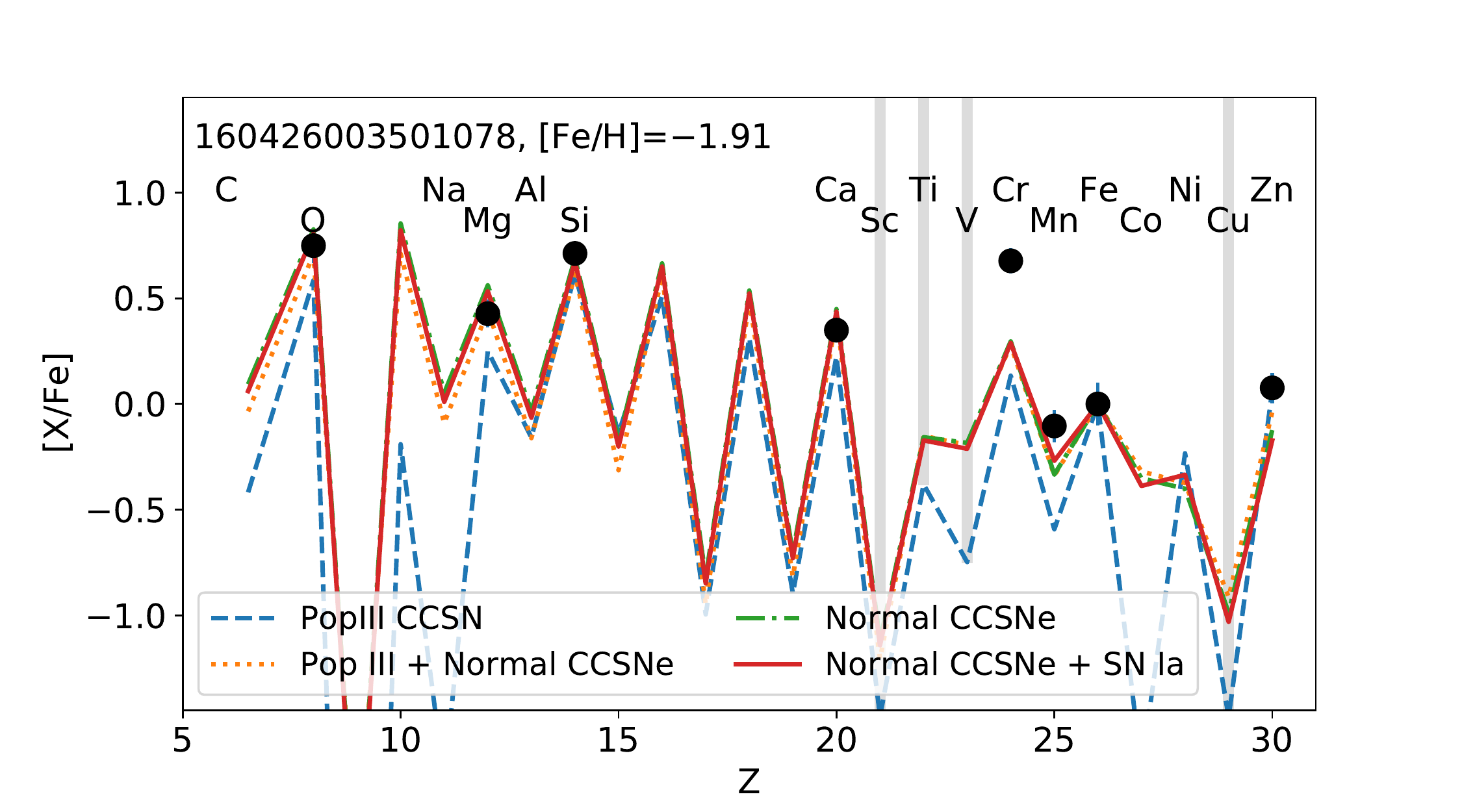}
    \caption{Summary of the 
    best-fit models obtained for the Models A (dashed line), 
    B (dotted line), C (dash-dotted line) and D (solid liinie) for the representative stars of the three OHS subgroups. The observed abundances are shown in black circles.}
    \label{fig:Compare_Models}
\end{figure}

Figure \ref{fig:Compare_Models} summarizes the best-fit 
model obtained for each of the Models A-D 
compared to the observed abundances 
for the representative stars of the three OHS subgroups. 

For both the high-$\alpha$ and low-$\alpha$ OHS shown in the 
top two panels of Figure \ref{fig:Compare_Models}, either 
the Pop~III + normal CCSNe (Model B; dotted line) 
or the normal CCSNe + SN~Ia (Model D; solid line) yields
provide a better fit than the other two models. 
The Pop~III CCSN model (Model A; dashed line) tends to significantly 
under-predict the abundance ratios of odd-atomic-numbered elements 
such as Al, or Cu. The normal 
CCSN model (Model C; dash-dotted line) 
over-predict the observed abundance ratios of 
elements from O to Si. 
In the case of the metal-poor OHS shown in the bottom panel, 
the difference among the
best-fit Models B-D is small since the yields of 
normal CCSN with low metallicity  dominate 
over the Pop III CCSNe or SN~Ia.

We note that a quantitative comparison among Models A-D is not straightforward because the four 
models employ different 
numbers of model parameters, which are not nessesarily 
independent. Therefore, the 
comparison of the reduced $\chi2$ values between different 
models should be viewed with caution. In the next subsections, we 
discuss validity of each model in terms of 
other constraints from simulations and observations.

\subsection{Pop~III CCSNe \label{sec:discussion_PopIII}}

Cosmological simulations including metal enrichment 
by early generations of massive stars generally predict that the metal pollution is 
patchy and thus pristine gas for the formation of Pop~III stars 
can survive over a wide range of cosmic time \citep[e.g.][]{maio10}. 
With cosmological hydrodynamical simulations 
over a volume of $(10{\rm ~Mpc}h^{-1})^3$, 
\citet{pallottini14} suggest that the formation of 
Pop~III stars from pristine gas can occur 
at least up to $z=4$ in regions far from star-forming galaxies and 
in low mass halos ($M_{\rm h}\lesssim ~10^8M_\odot$). 
In the following we examine the validity of the Pop~III CCSN 
enrichment scenario in terms of 
best-fit Pop~III CCSN models obtained in the previous section 
(Section \ref{sec:bestfitPopIII}) and in terms of 
semi-analytical models for chemical enrichment in 
the context of the hierarchical galaxy formation scenario 
(Section \ref{sec:simulation}).

\subsubsection{Best-fit Pop~III CCSN model parameters \label{sec:discussionPopIII}}

If the OHS are actually very old stars (e.g., $>12$ Gyrs), 
they could potentially be the first 
metal-enriched stars formed out of gas 
locally pre-enriched by Pop~III stars. 
As the result of comparing observed abundances  
with the Pop~III CCSN yield models, we find that the abundance patterns of 
the OHS are best explained by the Pop~III stars 
with progenitor masses of 15 or 25 $M_\odot$.

One difficulty of the Model A is that the 
ISM from which the OHS have formed should have been  pre-enriched with neutron capture elements, such as Y or 
Ba, both detected in the OHS. These elements 
are synthesized by both rapid (r-) and slow (s-) neutron capture 
processes, which generally require Fe seed nuclei. 
These processes, therefore, 
are not likely associated with 
Pop~III stars with strictly zero metal content (see \citealt{choplin20} for a 
possible channel to produce neutron-capture elements in 
a low-metallicity massive star). 
The abundances of the [Ba/H] and [Y/H] in the 
OHS are a several orders of magnitudes higher 
than typical EMP stars and an order of magnitude higher than 
the values observed in the r-rich ultra-faint dwarf galaxy 
\citep[e.g.,][]{ji16}.  Moreover, since the [Ba/Y] ratio 
of the OHS are close to the solar value, both r- and s-
process should have contributed unlike the pure r-process 
ratios reported by \citet{ji16}.

Another potential difficulty of the scenario 
is that the required mass of hydrogen to dilute ejected 
Fe to be compatible with the observed [Fe/H] value is 
extremely small. 
With an analytic prescription \citep[e.g.,][]{thornton98,tominaga07}, 
the swept up hydrogen mass by ejecta of a supernova with a 
given explosion energy of $E_{51}=1$ is estimated to be 
10$^5~M_\odot$ for the hydrogen number density, n=1--100 cm$^{-3}$. 
In contrast, the hydrogen dilution mass required to 
explain both [X/Fe] ratios and [Fe/H] is determined to be 
100--1000$~M_\odot$, which
is much smaller than the analytic prescription. This 
value is also not compatible with a new 
estimate of the minimum dilution mass 
by \citet{magg20}.
In those respect, it is unlikely that only a single Pop~III supernova 
dominate in producing metals in the OHS. 

In order to more robustly conclude on the possible 
chemical signature of Pop~III stars in the OHS, 
theoretical investigations on the possible distributions 
of metallicity of the first 
metal-enriched stars are nessesary \citep{karlsson13,smith15}. 
It has been proposed that various 
different mechanisms could have played a role in determining 
the condition at which the first metal-enriched stars form, 
potentially leading to the [Fe/H] spread. They includes  
(1) the properties of the Pop~III 
stellar systems such as the IMF, multiplicity, 
and the number of Pop~III stars per mini halos 
\citep{clark11,greif11,stacy14,susa14,hartwig18}, 
(2) hydrodynamical properties of Pop~III SN ejecta that 
could depend on the explosion energy and geometry 
\citep{joggerst09,tominaga09,ritter12}, 
(3) the properties of the interstellar medium 
to which energy and metals from Pop~III SNe are injected 
\citep{kitayama05,greif10,jeon14,chiaki18,tarumi20}, 
(4) metals and dust abundances that determine 
the efficiency of the formation of the next-generation 
low-mass stars \citep[e.g.,][]{omukai05,chiaki14,debennassuti14,hartwig19b} 
and (5) the redshift evolution of CMB and the properties of the host halos \citep{tumlinson07,smith09}. 
Predictions on their combined effects on the [Fe/H] 
spread among the first metal-enriched stars would provide 
useful insights into the best strategy for the 
upcoming Galactic Archaeology surveys.

\subsubsection{Can stellar ages help to select stars purely enriched by Pop~III stars?\label{sec:simulation}}
Based on the paradigm of hierarchical structure formation and incremental chemical enrichment over comic time, we expect a causal connection between the age and chemical composition of a star. 
Our finding of several old stars at [Fe/H]$>-1$, 
with abundance ratios similar to Pop~III CCSN 
yield models raises the question if we can use this insight to select interesting candidates for stellar archaeology based on an age selection.

To answer these questions, we use the semi-analytical model \textsc{a-sloth} (Ancient Stars and Local Observables by Tracing Haloes)\footnote{\url{http://www-utap.phys.s.u-tokyo.ac.jp/~hartwig/A-SLOTH}} based on \citet{hartwig18} and Magg et al. in prep. with an improved subgrid model for stochastic metal mixing in the first galaxies \citep{tarumi20}. More technical details of \textsc{a~sloth} can be found in the corresponding references, and we briefly summarize its main features here. 

\begin{figure}
    \centering
    \includegraphics[width=8cm]{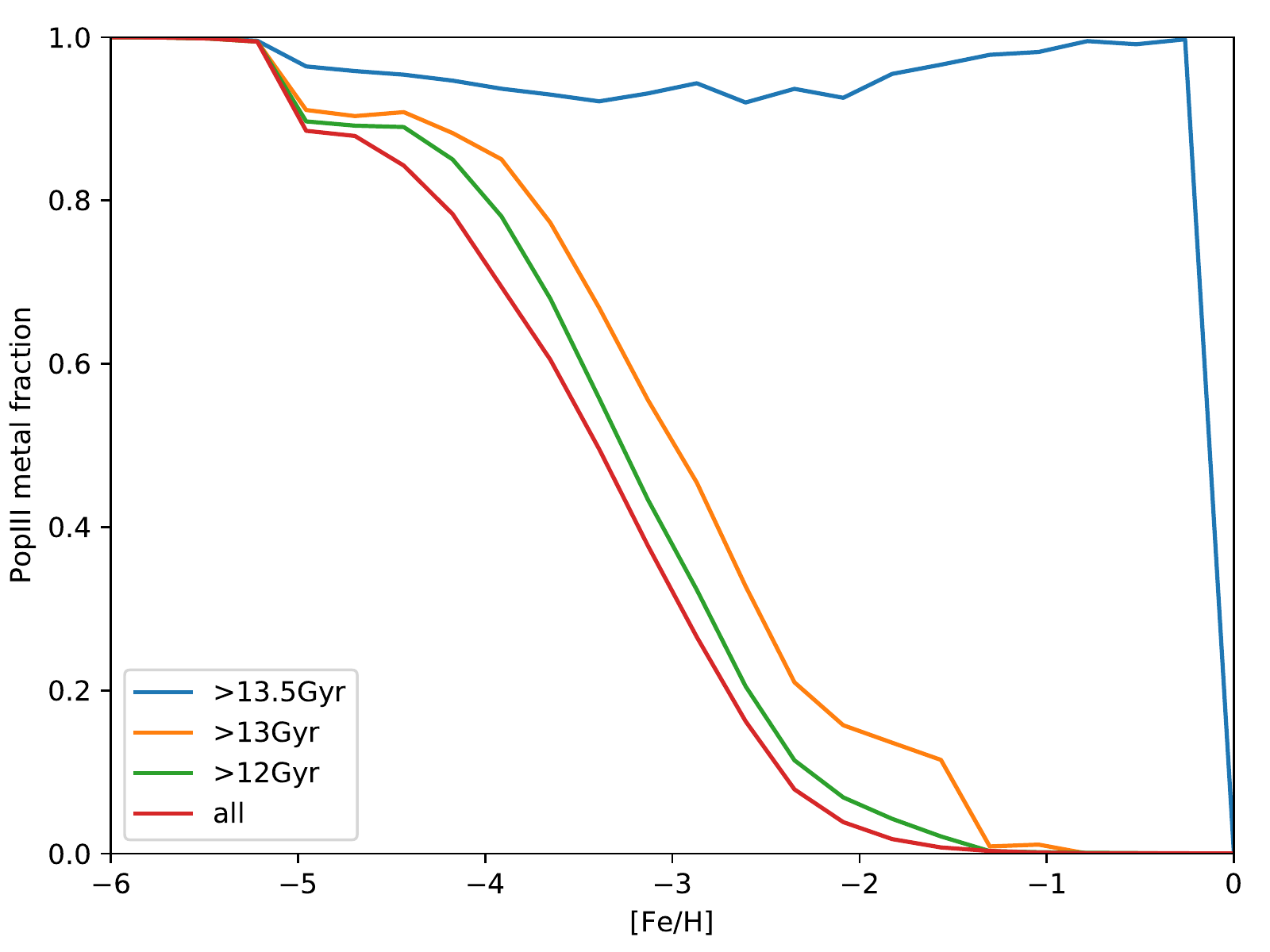}
    \caption{Simulated fraction of metals in MW stars that come from Pop~III SNe as a function of stellar metallicities. The different lines illustrate how an age cut helps to select interesting candidates that are dominated by metals from Pop~III SNe. A theoretical age selection of $>13.5$\,Gyr could help to select such Pop~III-dominated stars at metallicities up to [Fe/H]$ \sim 0$.}
    \label{fig:FeHFrac}
\end{figure}

On top of 30 Milky Way-like dark matter merger trees from the Caterpillar simulation \citep{griffen16}, we model Pop~III star formation by following chemical, radiative, and mechanical feedback. This allows us to analyse the chemical enrichment history of the Milky Way with detailed abundance and age information of stars that end up in the Milky Way halo at $z=0$. In our fiducial model \citep[calibrated against the metallicity distribution function, see ][]{tarumi20}, we use a Pop~III IMF from 2-180\,M$_\odot$ with a slope of $\mathrm{d}N/\mathrm{d}M=-0.5$, and a Pop~III star formation efficiency of $1\%$. With this model, we can predict what fraction of metals in a star comes from Pop~III SNe or from later generations of stars. The results as a function of metallicity are illustrated in Figure \ref{fig:FeHFrac}.
The overall trend for all stars (red line) shows that the average fraction of metals from Pop~III SNe decreases monotonically with metallicity. The contribution of metals from Pop~III SNe to stars with [Fe/H]$>-2$ is only $<10\%$. The semi-analytical model predicts that metals from Pop~III stars are insignificant for the OHS with [Fe/H]$>-1$. However, if we could select old stars with $>13.5$\,Gyr, this picture changes dramatically. The metal content of such old stars is dominated ($>80\%$) by Pop~III SNe at all metallicities up to solar. Although challenging in reality, such an age pre-selection could be very valuable to pre-select interesting candidates for stellar archaeology at [Fe/H]$>-3$.

This plot also shows another property of general interest: the fractional Pop~III contribution as a function of metallicity. For EMP stars ([Fe/H]$<-3$), at least $50\%$ of their metal mass comes from Pop~III SNe. This value increases to $\gtrsim 80\%$ at [Fe/H]$<-4$.

\begin{figure}
    \centering
    \includegraphics[width=8cm]{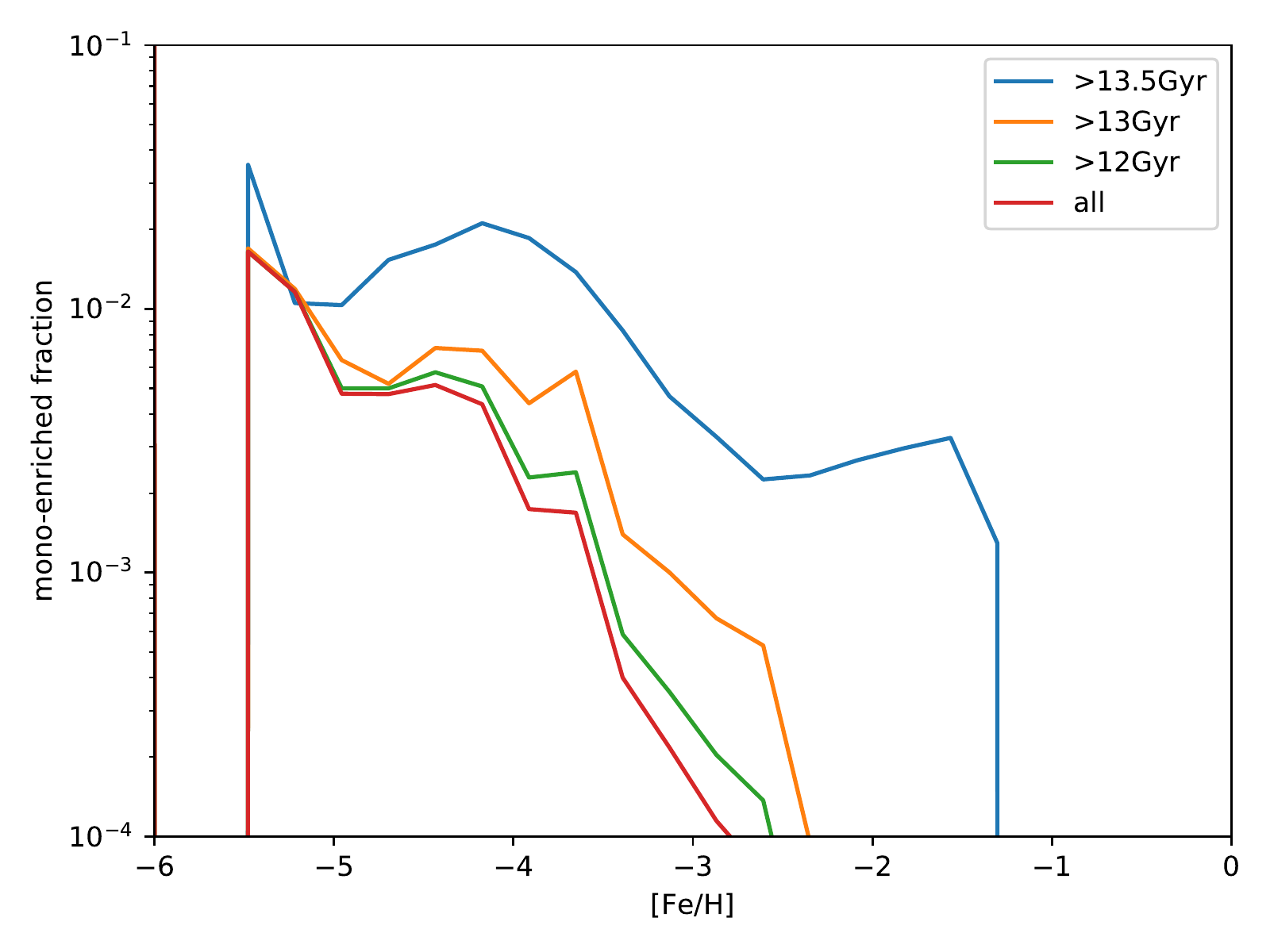}
    \caption{Fraction of stars that are enriched by only one Pop~III SN as a function of metallicity. The lines represent different age cuts, which may allow to pre-select interesting candidates. While an age-cut increases the fraction of mono-enriched stars at all metallicities, there are no mono-enriched stars at [Fe/H]$>-1$ in our model.}
    \label{fig:FeHMono}
\end{figure}

Another interesting related question is if we can identify mono-enriched stars at higher metallicities with the help of an age selection. We show the mono-enriched fraction as a function of metallicity in Figure \ref{fig:FeHMono}.
An age-cut can help to increase the fraction of mono-enriched stars in the sample, but the requirements of age precision are beyond current isochrone-based estimates. Moreover, the overall numbers of mono-enrichment depend strongly on the assumed Pop~III star formation efficiency, which is only weakly constrained. In summary, and age-cut is a valuable tool to select interesting, informative candidates for stellar archaeology, but the precision of the age estimate needs to be improved for this method to be reliable.

\subsection{Pop~III and Normal CCSNe} 

We next discuss whether the observed abundances in 
the OHS imply the scenario that both Pop~III and 
normal CCSNe contribute to the metal enrichment (Model B). 
In what condition this scenario is realized is not clear 
because of the complexity in physical mechanisms that determine 
the degree of homogeneity in star forming environment 
in the early Universe. A motivation of this scenario is that, 
at redshift $z>3$, when a sizable fraction of Milky Way halo stars 
formed, it may be possible that, 
the systems is mainly chemically enriched by normal CCSNe, 
while there is remaining pristine gas, which could 
form Pop~III CCSN progenitors.

We have found that, the 
contributions from both Pop~III and normal CCSNe reasonably 
well explain observed elemental abundances in the OHS. The combined yields simultaneously 
explain the abundance ratios for both the intermediate-mass 
elements (Na-Si) and the iron-peak elements (Cr-Zn) 
(right panels of Figure \ref{fig:PopIII_abupattern}). 

Similar to the Model A, nearly solar [Y/Fe], 
and [Ba/Fe]  ratios in the OHS remain difficult to 
explain, which requires a pre-enrichment of 
both r- and s-process elements to the ISM prior to 
the Pop~III CCSN.  Validity of this scenario is, therefore, depends on the possible sources of the neutron-capture 
elements in the cosmic epoch when Pop~III stars 
was still contributing to the chemical enrichment.

\subsection{Normal CCSNe}

Under the assumption that yields of normal CCSNe dominate
the metal abundances in the OHS (Model C), 
none of the yield models we have taken into account explain the data better than the other three models.
For the high-$\alpha$ and low-$\alpha$ OHS with [Fe/H]$>-1.5$, 
the predicted abundance patterns of elements from O to Si relative to Fe are much higher than observed values unless 
unusually steep IMF slopes such as $\alpha_{\rm IMF}>3$ are assumed. 
This result is robust against change in 
the progenitor metallicity ($Z_{\rm CC}$). 
 We therefore conclude that 
the yields of normal CCSNe alone is not likely as the 
origin of the metal abundances in the OHS, 
unless the IMF is extremely bottom heavy.

Observations of high-redshift galaxies or different
Galactic regions have reported a possible 
signature of variations in the IMF 
with cosmic time or with local environment \citep{bastian10}. In particular, 
recent studies have provided evidences that the 
IMF is top-heavy for environments with high 
star formation rates \citep[][and reference therein]{cowley19}. 
The values of $\alpha_{\rm IMF}$ to better explain 
the OHS's abundances, on the other hand, imply
a significantly steeper or a bottom-heavy 
IMF for the normal CCSN progenitors, which is not 
motivated by any other observations \citep{bastian10}.

\begin{comment}
We note that the Ni yields of CCSNe 
can be subject to theoretical uncertainties.  
In core-collapse supernovae Ni is produced in the 
complete Si burning during the explosive nucleosynthesis. 
The predicted yield of Ni thus strongly 
depend on the assumption 
about the mass cut. We therefore need to 
keep in mind that the results can be affected 
by this assumption about the mass cut for the CCSNe. 
\end{comment}

\subsection{SNe~Ia} 

The cumulative contribution of SNe Ia to metals 
in the present Universe has been studied through  
the Solar chemical composition 
\citep[e.g.,][]{tsujimoto95} or the abundances in the
intra-cluster medium (ICM) in nearby clusters of galaxies 
\citep{matsushita03,deplaa07,simionescu15,mernier16}. 
In one of the latest studies on this topic, \citet{simionescu19} employed recent CCSN and SN~Ia yield calculations 
to explain the observed abundance ratios 11 different chemical 
elements detected in the core of the Perseus cluster of galaxies 
from high-resolution X-ray spectroscopy. Depending on 
the yield models used, they find that 
13--40 \% of SNe that have contributed to the 
metal enrichment are SN Ia and that 9--36 \% of all SNe~Ia 
are associated with near-$M_{\rm Ch}$ progenitors.

 How the SN Ia rate changes as a function of cosmic time or of environments
remains elusive because there is no consensus about 
the progenitor systems and the mechanisms which make 
the system to finally explode. 
The observed lack of evolution of the Fe content in the ICM of clusters of galaxies out to redshift $\sim2$ \citep{mcdonald16,mantz17,liu20,mantz20} suggests that metal enrichment by SN~Ia was already important early during cosmic history. This is further confirmed by measurements of the metal abundances in the outskirts of nearby galaxy clusters \citep{werner13,urban17}; the remarkably uniform distribution of Fe over large spatial scales, and the small cluster-to-cluster scatter, suggest that the ICM was enriched more than 10 billion years ago, before these clusters developed a strongly stratified entropy gradient that would prevent the efficient mixing of metals.

The SN~Ia rate evolution at high redshifts has also been addressed through the characteristic delay time or the delay time distribution for SN Ia 
\citep[e.g.,][]{hopkins06,totani08,hachisu08,maoz14}. 
Based on the observed cosmic star formation rate density evolution, \citet{hopkins06} suggest a characteristic delay time of $t\sim 3$ Gyrs, with no strong evidence of a ``prompt'' 
component (i.e., no time delay). On the other hand, other studies suggest that SN~Ia can explode sooner after a starburst event. \citet{totani08} have found that a delay time 
distribution of the form $\propto t^{-1}$ at delays $>1$ Gyrs best explains their observations of SN~Ia rates for a sample of elliptical galaxies, which is generally consistent with the DD scenario (see, however, \citealt{hachisu08} for the 
explanation of the $t^{-1}$ delay time distribution with the SD scenario). \citet{maoz12} find a continuous delay time distribution, with significant detections of prompt ($<0.4$ Gyr), intermediate ($0.4-2.4$ Gyr), and delayed ($>2.4$ Gyr) explosions.

If  the ages of the OHS are accurately greater than 12 Gyrs, they provide additional 
insights into the SN~Ia rates at high redshifts. 
Fitting the abundance ratios of $\alpha$ and Fe-peak elements 
simultaneously helps alleviating 
the degeneracy between the SN Ia fraction and the IMF 
slope of CCSN progenitors. 
We find that 
the model in which 4-6\% of all the metal-enriching SNe are 
SN~Ia best explains the abundances of the OHS with [Fe/H]$>-1.5$. 
The smaller relative SN Ia contribution requires 
the IMF of the CCSN progenitors to be 
unusually steep ($>3$).
The Type Ia supernovae are so far the only channel for synthesizing
Mn consistent with the solar composition
\citep{Seitenzahl2013, Nomoto2017SNIa}. 
The observed [Mn/Fe] ratios in the OHS, therefore, provide 
strong hints on the previous contamination by SNe~Ia.

Although the SN~Ia enrichment in the first few billion 
years of the Universe 
remains elusive \citep{maoz14,hopkins06}, the results presented here
are in line with the early chemical enrichment inferred 
from metal abundance distributions in nearby and high-redshift galaxy clusters \citep[e.g.,][]{mantz20}. 
It is, however, not well established whether near-$M_{\rm Ch}$ white dwarfs or sub-$M_{\rm Ch}$ 
white dwarfs 
are the dominant SNe~Ia progenitors 
at high redshifts. The ICM abundances in galaxy clusters 
hint at a certain contribution from sub-$M_{\rm Ch}$ SN~Ia enrichments \citep{simionescu19}. 
Chemical abundance ratios in the stars of ancient dwarf Milky Way satellite galaxies also support contributions of sub-$M_{\rm Ch}$ SN~Ia \citep{kirby19,kobayashi20b}.

\subsection{Contributions from other sources \label{sec:othersources}} 

 In this section we discuss 
the abundances of elements that are potentially 
affected by nucleosynthetic sources other than Pop~III/normal CCSNe 
or SN~Ia.

\subsubsection{Y and Ba} 

For the solar system material, more than 80 \% of Y and Ba 
are attributed to main s-process in 
AGB stars \citep{arlandini99,prantzos20}. In the early 
Galaxy, stellar winds from 
fast-rotating massive stars \citep{meynet06,hirschi07,ekstrom08,yoon12} are predicted to 
have significant contribution to the metal enrichment, 
including s-process elements like Y or Ba \citep{pignatari08,chiappini11,frischknecht16,choplin18,limongi18,prantzos18}. In addition, Ba is likely synthesized by 
the main $r$-process in the early Universe, 
whose major astrophysical site is 
subject to debate \citep[e.g.,][]{cowan19}.  In addition to the main $r$-process, 
other sources are required for the production of Y 
to be compatible with observed abundance patterns in 
EMP stars \citep[e.g.,][]{francois07}.

\begin{figure}
    \centering
    \includegraphics[width=8cm]{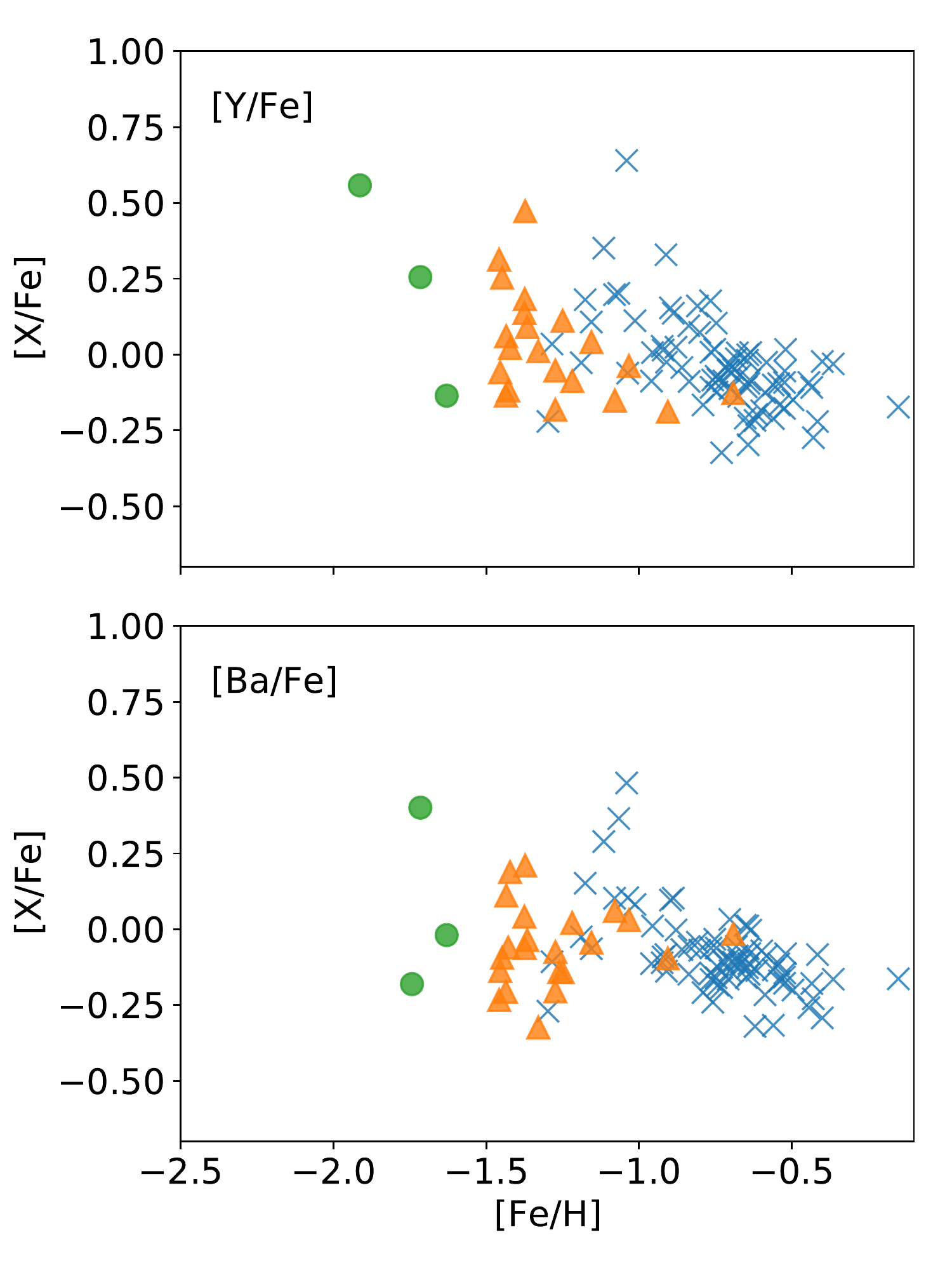}
    \caption{[Y/Fe] and [Ba/Fe] ratios of the three OHS subgroups. 
    Symbols are the same as in Figure \ref{fig:xfe_feh}.}
    \label{fig:ba_y}
\end{figure}

Figure \ref{fig:ba_y} shows the [Y/Fe] and [Ba/Fe] abundance 
ratios for the three OHS subgroups. The low-$\alpha$ OHS 
show lower [Y/Fe] than the high-alpha OHS at [Fe/H]$\sim -1$ 
as reported by preceding studies \citep{ishigaki13,fishlock17,matsuno20}. 
On the contrary, 
the [Ba/Fe] ratios of the low-$\alpha$ OHS are indistinguishable from 
the high-$\alpha$ OHS. The metal-poor OHS subgroup is characterized 
by large scatter in both the [Y/Fe] and [Ba/Fe] ratios. 
The diversity of these neutron-capture 
elemental abundances hints at the environmental dependence in the 
s-process enrichment among the oldest nearby halo stars.
In order to clarify 
which of the proposed sites are responsible for producing each of Y and Ba 
in the OHS, abundance determinations of C and N as well as 
other neutron-capture elements are necessary.

\subsubsection{Zn} 

Another sources, that could have contributed 
to the observed abundances in 
OHS are nucleosynthetic products of stars 
with masses $\sim 8-10 ~M_{\odot}$, which end their life as 
electron-capture supernovae \citep{nomoto84a,nomoto87}. 
Electron capture supernovae are triggered by 
electron capture on $^{20}$Ne when the mass of the 
electron-degenerate O-Ne-Mg core of $\sim 8-10~M_{\odot}$ stars becomes near the Chandrasekhar mass limit 
($\sim 1.4 M_\odot$) so that the
  central density gets close to the threshold density
  \citep{nomoto84a,nomoto87,Nomoto2017ECSN,Leung2019PASA,jones19a}. 
The contribution to GCE models are thought to be negligible, except for neutron-capture elements  \citep{kobayashi20a}. 
The evolution of the progenitor with O-Ne-Mg core based
 on most recent microphysics model \citep{Suzuki2019} suggest that
 electron-capture supernova is triggered at the central density of log $\rho_c$(g cm$^{-3}) > 10.01$ in the majority of parameter space on
convection, etc., which is very likely to lead to collapse to form
 neutron stars \citep{Zha2019,Leung2020}.

An alternative channel for synthesizing matter
with a high Zn/Fe ratio is the collapsar \citep{Tsuruta2018}. The high velocity jet triggered
by the central rotating black hole formed by the
core in a massive star provides the necessary shock
heating. The thermal energy creates the necessary high
entropy zone during the alpha-rich freezeout burning
\citep{Maeda2002,matsuba2004}.

To test 
whether the possible contribution of Zn other than CCSNe or 
SN~Ia could affect our conclusion 
on the relative CCSNe and SNIa fractions, we repeat the fitting  
for the Model D with $f_{\rm Ch}=1.0$, 
considering the model abundance of Zn 
as a lower limit. 
The results on the best-fit model parameters 
do not change significantly, since all the models shown in 
Fig \ref{fig:ccsn_Ia}  
under-predict [Zn/Fe]. This indicates that 
the possible contribution from additional Zn sources 
does not affect the main conclusion in the present analysis.

\subsection{Relation to known Galactic stellar populations}
\label{sec:MWformation}

The OHS analyzed in this study have been selected 
to have orbital kinematics 
compatible with the stellar halo population of the Galaxy (Section 
\ref{sec:selectkin}). 
The origin of the stellar halo has been debated for many decades
\citep[e.g.,][]{eggen62,searle78,chiba00,carollo07}, 
and it is still being actively discussed \citep[e.g.,][]{helmi20}. In particular, astrometric data 
from {\it Gaia} combined with ground-based massive 
photometric and spectroscopic surveys have made a major breakthrough 
in this discussion by the discovery of a clear evidence of 
a merger at the early Galactic formation epoch ($z\sim 1-2$) 
\citep[e.g.,][]{belokurov18,helmi18}. 
The debris stars of this merger event, called ``Gaia-Enceladus-Sausage (GES)'', are found to constitute a large fraction of the local stellar halo population \citep[e.g.,][]{dimatteo19,naidu20}. The discovery of 
GES is in line with the hierarchical formation of the 
Galaxy as predicted by cosmological simulations 
\citep{bullock05,font06,delucia08,cooper10}. 
The majority of remaining local halo stars  
are found to be on prograde orbits and have 
metallicities similar to the thick disk population
\citep{haywood18,dimatteo19,bonaca20,belokurov20}. 
These characteristic kinematics and metallicities are 
compatible with the population formed in-situ, presumably 
in the old disk or bulge, part of which 
have dynamically heated by accretion events \citep{zolotov10,purcell10,tissera13}. The dynamical 
heating is thought to be largely associated with 
the GES merger event, which is predicted to have occurred 
 8-10 Gyrs ago \citep{helmi18,mackereth19,gallart19,belokurov20,grand20}. 
 
The OHS provides a unique opportunity to observationally 
constrain the early Galactic environment, 
prior to the merger of GES, 
thanks to the detailed elemental abundance measurements 
from the GALAH catalogs \citep[see also][]{fernandez-alvar18}. 
The high-$\alpha$ OHS includes stars with [Fe/H] as high as $\sim -0.3$. We have shown that, regardless of the assumptions about 
the IMF of normal CCSN progenitors, the 
observed abundances of the high-$\alpha$ OHS
are best explained by yields of the normal CCSNe with 
metallicities as high as $Z_{\rm cc}/Z_{\odot}\sim 0.1-0.3$, 
suggesting that the 
enrichment by the normal CCSNe should have been very 
efficient and have occurred on short timescales. At the same time,  
the observed abundances of the high-$\alpha$ OHS are 
better explained with a certain (up to 10 \%) contribution  
of SN~Ia. This implies that SN~Ia, which produce $\sim 10$ 
times more Fe than a normal CCSN, have played a role 
in the chemical enrichment in the earliest epoch of the Galaxy formation. 
The low-$\alpha$ OHS with [Fe/H] in the range from $-1.5$ 
up to $-0.7$ includes stars showing a sign of larger 
contribution from SN~Ia (up to 20 \%). 
On the other hand, the metal-poor OHS 
with [Fe/H]$<-1.5$ do not exhibit strong evidence of 
the SN~Ia nucleosynthetic pattern (Figure \ref{fig:hist_CCSN_Ia}). 
The wide range of [Fe/H] as well as [X/Fe] ratios among the OHS 
implies diversity 
in metal-enrichment sources available in the early Galactic 
environment.

\subsection{Future prospects for age estimation}
\label{sec:age_prospects}

This study highlights the importance 
of more accurate age estimates for the old (age $>12$ Gyrs) halo stars to 
robustly interpret
their elemental abundances in terms of metal enrichment 
sources in the early Universe. 
Indeed, stellar age dating has been one of the fundamental 
elements to make constraints on the Galactic chemical and dynamical 
evolution \citep[e.g.,]{edvardsson93,haywood13,schuster12}.
It has been, however, challenging to 
  obtain precision ages for a large statistical 
  sample of the stars belonging to the Milky Way halo, which are 
  very rare in the solar neighborhood. 
As detailed in 
Section \ref{sec:simulation} and shown in 
Figure \ref{fig:FeHFrac}, the models predict that the 
age cut of $>13$ Gyrs is required to select field stars 
with [Fe/H]$\gtrsim -2$ that
likely retain chemical signatures of Pop~III stars. None of the 
current samples of stars satisfies this criterion and the age 
errors up to a few Gyrs for the current OHS sample 
do not allow for a clean 
separation of such extremely old stars. 

Recent and on-going 
asteroseismology space missions such as 
{\it TESS}\citep{ricker15} and {\it K2} \citep{hawell14} will provide accurate stellar 
mass estimates with a precision of a few percent, 
which are crucial for identifying potentially oldest 
stars in the solar-neighborhood. 
Using these 
asteroseismology data as a training set, 
data-driven approaches to estimate stellar masses 
(and thus ages) from 
spectroscopic observations alone 
became a powerful tool, allowing 
for the age estimates for distant halo stars \citep[e.g.,][]{chaplin13,martig16,ho17,wu19,das19}. 
Planned large spectroscopic surveys including 
WEAVE \citep{bonifacio16}, 4MOST \citep{dejong19}, DESI \citep{desi16}, Milky Way Mapper \citep{kollmeier17}, or PFS \citep{takada14} 
combined with asteroseismic data will 
be promising to deliver stellar ages for a large volume in the stellar halo.

Nucleocosmochronometry is another 
technique to accurately estimate ages 
with a precision of less than $10$ percent \citep{soderblom10}, 
although it is currently not feasible to 
build a large sample with this measurement. 
Extremely large telescope such as GMT, ELT, 
and TMT would be needed to carry out the 
nucleocosmochronometry analysis for 
well-selected candidates of old halo stars.

\section{Conclusions} 
\label{sec:conclusion}

 We have investigated the origin of metals in 
nearby old Milky Way halo stars selected 
from GALAH DR3 \citep{buder21} and Gaia EDR3 \citep{lindegren20}. 
Based on stellar parameters ($T_{\rm eff}$, $\log g$ and [Fe/H]) 
and astrometric data (parallax and proper motion) 
provided by these 
catalogs, main-sequence turn-off stars 
with estimated ages greater than 12 Gyrs and 
with halo-like kinematics (``OHS'') are selected 
as candidates of stars belong to the old stellar 
population in the Solar neighborhood
with homogeneous detailed elemental 
abundance measurements. 

 We have tested different hypotheses about 
the sources of the metals in the 
high-$\alpha$, low-$\alpha$ and metal-poor OHS subgroups 
by comparing their observed abundance patterns ([X/Fe] versus 
atomic number) with the 
Pop~III CCSNe, normal CCSNe and SN Ia yield models.
The main results can be summarized as follows: 

\begin{enumerate}
    \item  Pop~III CCSN yields with 15--25$~M_\odot$ reasonably 
    well explain the observed abundance ratios ([X/Fe]) in the OHS (Model A; Section \ref{sec:bestfitPopIII}). The contributions from both Pop~III and normal CCSNe 
    simultaneously reproduce observed abundances of O-Si 
    and the Fe-peak elements (Model B; Section \ref{sec:PopIIICCSNe}). 
     However, significant contribution from 
    a single Pop~III CCSN, assumed in 
    both Model A and B, suffers from difficulties in explaining 
 the relatively high [Fe/H] for the OHS, 
 which requires an extremely 
    small mass of H that dilutes Fe (Section \ref{sec:discussionPopIII}). 
    Based on the semi-analytical model, we have also 
    shown that the contribution of Pop~III CCSN to the 
    metal content is insignificant  
   for the OHS with [Fe/H] as high as $\sim -1$ (Section \ref{sec:simulation}).
   
    \item Normal CCSNe yields with a characteristic 
    metallicity averaged over the initial mass function of the 
    power-low form has difficulty in explaining the [X/Fe] 
    ratios among O-Si, as well as some of the Fe-peak elements 
    unless an extremely bottom-heavy IMF is assumed 
    (Model C; Section \ref{sec:resultsNormalCCSNe}).
    
    \item Contributions from both normal CCSNe and SNe Ia 
    (Model D; Section \ref{sec:resultsCCSNeSNIa}) simultaneously explain observed elemental abundance patterns 
    of $\alpha$ and Fe-peak elements of the high-$\alpha$ and 
    low-$\alpha$ OHS subgroups with [Fe/H]$>-1.5$. 
    In this model, 
    the observed abundance patterns of the 
    high-$\alpha$ OHS subgroup  
    are best explained by sets of 
    model parameters where up to 10--20\% of all metal-enriching supernovae are 
    SNe~Ia. A higher contribution 
    from SN Ia up to 27 \% best explain 
    the abundances of low-$\alpha$ subgroup. 
    This fraction also depends on the assumed 
    fraction of near-$M_{\rm Ch}$ white dwarf 
    progenitors among all 
    the SN~Ia progenitors. For the OHS analyzed 
    in this study, 
     $~50\%$ contribution from a near-$M_{\rm Ch}$ progenitors best explains the observed abundances.

 In terms of the yield models and the model parameters
 we have took into account, 
 the last scenario provides the best description of the abundance ratios at least for stars with [Fe/H]$>-1.5$ (Section \ref{sec:discussionComparison}). The inferred diversity 
 in metal enrichment sources among the OHS 
 has implications on the formation environment of
 the oldest stellar population in the Solar neighborhood, 
 while the exact timing of the enrichment 
 depends on stellar absolute ages, which 
 is currently uncertain. We also note that none of the models we have tested are a perfect match to the data, therefore improvements of the observational constraints and of the theoretical yields are necessary to draw robust conclusions.

\end{enumerate}

 Future spectroscopic, astrometric and astroseismology surveys will deliver kinematics 
and detailed elemental abundances as well as age estimates 
for the Galactic stellar populations with significantly improved statistics. Combined with improvements in 
theoretical yield calculations and cosmological simulations of the early Galactic chemical enrichment, 
these data provide us with 
an opportunity to more robustly constrain the nucleosynthetic 
origin of individual stars and thus 
help investigating the chemical and dynamical 
history of the early Galaxy in unprecedented 
details.

\section*{Acknowledgements}

The authors thank the anonymous referee for useful comments, which significantly improved this manuscript.  We are very grateful to S. Sharma for helpful assistance on stellar ages from GALAH DR3.  
This work has been supported by the World Premier International
Research Center Initiative (WPI Initiative), MEXT, Japan.
This work has also been supported by JSPS KAKENHI Grant Numbers 17K14249, 18H05437, 20H05855 (M. N. I.), 19K23437, 20K14464 (T. H.), JP17K05382 and JP20K04024 (K. N.), JP21H04499 (M.N.I, N.T., and K.N.).
S.C.L. acknowledges support by NASA grants HST-AR-15021.001-A and 80NSSC18K101.
A. S. is supported by the Women In Science Excel (WISE) programme of the Netherlands Organisation for Scientific Research (NWO), and acknowledges the World Premier Research Center Initiative (WPI) and the Kavli IPMU for the continued hospitality. SRON Netherlands Institute for Space Research is supported financially by NWO.
C. K. acknowledges funding from the UK Science and Technology Facility Council (STFC) through grant ST/M000958/1 \& ST/R000905/1.
This work was supported in part by the National Science Foundation under Grant No. OISE-1927130 (IReNA). 
This work made use of the Third Data Release of the GALAH Survey \citep{buder21}. The GALAH Survey is based on data acquired through the Australian Astronomical Observatory, under programs: A/2013B/13 (The GALAH pilot survey); A/2014A/25, A/2015A/19, A2017A/18 (The GALAH survey phase 1); A2018A/18 (Open clusters with HERMES); A2019A/1 (Hierarchical star formation in Ori OB1); A2019A/15 (The GALAH survey phase 2); A/2015B/19, A/2016A/22, A/2016B/10, A/2017B/16, A/2018B/15 (The HERMES-TESS program); and A/2015A/3, A/2015B/1, A/2015B/19, A/2016A/22, A/2016B/12, A/2017A/14 (The HERMES K2-follow-up program). We acknowledge the traditional owners of the land on which the AAT stands, the Gamilaraay people, and pay our respects to elders past and present. This paper includes data that has been provided by AAO Data Central (datacentral.aao.gov.au).
This work has made use of data from the European Space Agency (ESA) mission
{\it Gaia} (\url{https://www.cosmos.esa.int/gaia}), processed by the {\it Gaia}
Data Processing and Analysis Consortium (DPAC,
\url{https://www.cosmos.esa.int/web/gaia/dpac/consortium}). Funding for the DPAC
has been provided by national institutions, in particular the institutions
participating in the {\it Gaia} Multilateral Agreement. 
This research made use of Astropy,\footnote{\url{http://www.astropy.org}} a community-developed core Python package for Astronomy \citep{astropy:2013, astropy:2018}, galpy \citep{bovy15}, NumPy \citep{van2011numpy}, Matplotlib \citep{Hunter:2007}, SciPy \citep{2020SciPy-NMeth}, 
Pandas \citep{McKinney_2010,McKinney_2011}, and PyMC3 \citep{salvatier16}.

\noindent
{\it Data availability}: The data underlying this article will be shared on request to the corresponding author.

%%%%%%%%%%%%%%%%%%%%%%%%%%%%%%%%%%%%%%%%%%%%%%%%%%


\bibliographystyle{mnras} % style aa.bst
\bibliography{ref} % your references Yourfile.bib

%%%%%%%%%%%%%%%%%%%%%%%%%%%%%%%%%%%%%%%%%%%%%%%%%%

%%%%%%%%%%%%%%%%% APPENDICES %%%%%%%%%%%%%%%%%%%%%

\appendix

\section{Estimated model parameters }
Tables \ref{tab:modelA_results}--\ref{tab:modelD_results} summarize the parameter estimates of the Models A--D for the OHS analyzed in this paper. The full tables in a machine readable format are available 
in the electronic edition of the journal. Only a portion is shown below as a guide.

\begin{table*}
    \centering
    \begin{tabular}{lrlrrrrrrrr}
\hline
    GALAH DR3 ID &  [Fe/H] & OHS subgroup &    $M$ & $E_{51}$ & $M_{\rm mix}$ & $\log f_{\rm ej}$ & $M_{\rm H}$ & $M_{\rm Ni}$ & $\chi^2$ & DoF \\
\hline
 131217003901110 & $-0.70$ &   high-alpha & $25.0$ &   $10.0$ &         $4.0$ &            $-0.7$ &  $3.20e+02$ &   $1.36e-01$ & $186.03$ & $6$ \\
 140116004302064 & $-1.63$ &   metal-poor & $25.0$ &    $1.0$ &         $3.1$ &            $-0.3$ &  $2.91e+03$ &   $1.35e-01$ &  $21.97$ & $2$ \\
 140303001002016 & $-0.91$ &   high-alpha & $15.0$ &    $1.0$ &         $1.6$ &            $-0.6$ &  $1.32e+02$ &   $3.55e-02$ &  $63.45$ & $7$ \\
 140412001201275 & $-0.76$ &   high-alpha & $15.0$ &    $1.0$ &         $1.6$ &            $-0.4$ &  $1.79e+02$ &   $5.51e-02$ &  $66.93$ & $5$ \\
 140413002701263 & $-0.58$ &   high-alpha & $25.0$ &   $10.0$ &         $4.0$ &            $-0.8$ &  $1.97e+02$ &   $1.08e-01$ &  $95.51$ & $5$ \\
\hline
\end{tabular}

    \caption{The summary of the best-fit yield models for Model A. The columns 4-8 show the best-fit 
parameters; Pop~III progenitor mass ($M$ [$M_\odot$]), 
explosion energy ($E$ [$10^{51}$ erg]),size of the mixing zone in the progenitor star's mass coordinate 
($M_{\rm mix}$), the ejected fraction ($\log f_{\rm ej}$), hydrogen 
dilution mass ($M_{\rm H}$ [$M_{\odot}$]).
%and the number fraction of core-collapse supernovae %relative to the total number of the Pop~III and CC %supernovae. 
The column 9 shows the ejected mass of $^{56}{\rm Ni}$ ($M_{\odot}$) of the best-fit yield. 
The $\chi^2$ and the degree of freedom (DoF) are given in columns 10-11.  }
    \label{tab:modelA_results}
\end{table*}

\begin{table*}
    \centering
    \begin{tabular}{lrlrrrrrrrrr}
\hline
    GALAH DR3 ID &  [Fe/H] & OHS subgroup &    $M$ & $E_{51}$ & $M_{\rm mix}$ & $\log f_{\rm ej}$ & $M_{\rm H}$ & $M_{\rm Ni}$ & $f_{\rm CC}$ & $\chi^2$ & DoF \\
\hline
 131217003901110 & $-0.70$ &   high-alpha & $25.0$ &   $10.0$ &         $2.1$ &            $-0.1$ &  $1.71e+03$ &   $6.06e-01$ &       $0.80$ & $125.38$ & $5$ \\
 140116004302064 & $-1.63$ &   metal-poor & $25.0$ &   $10.0$ &         $2.1$ &            $-0.2$ &  $1.32e+04$ &   $5.47e-01$ &       $0.80$ &  $10.82$ & $1$ \\
 140303001002016 & $-0.91$ &   high-alpha & $15.0$ &    $1.0$ &         $1.6$ &            $-0.7$ &  $1.31e+02$ &   $2.86e-02$ &       $0.20$ &  $55.95$ & $6$ \\
 140412001201275 & $-0.76$ &   high-alpha & $25.0$ &   $10.0$ &         $2.1$ &            $-0.1$ &  $1.97e+03$ &   $6.06e-01$ &       $0.80$ &  $57.97$ & $4$ \\
 140413002701263 & $-0.58$ &   high-alpha & $15.0$ &    $1.0$ &         $2.1$ &            $-0.1$ &  $2.31e+02$ &   $1.08e-01$ &       $0.50$ &  $65.69$ & $4$ \\
\hline
\end{tabular}

    \caption{The summary of the best-fit yield models for Model B. The columns 4-8 are the same as in 
    Table \ref{tab:modelA_results} and the column 9 
    shows the fraction of normal CCSNe. 
%and the number fraction of core-collapse supernovae %relative to the total number of the Pop~III and CC %supernovae. 
The column 10 shows the ejected mass of $^{56}{\rm Ni}$ ($M_{\odot}$) of the best-fit yield. 
The $\chi^2$ and the degree of freedom (DoF) are given in columns 10-11. }
    \label{tab:modelB_results}
\end{table*}

\begin{table*}
    \centering
    \begin{tabular}{lrlrrrrrr}
\hline
    GALAH DR3 ID &  [Fe/H] & OHS subgroup & $\alpha_{\rm IMF}$ & $\alpha_{\rm IMF, min}$ & $\alpha_{\rm IMF, max}$ & $Z_{\rm CC}$ & $Z_{\rm CC, min}$ & $Z_{\rm CC, max}$ \\
\hline
 160330103301048 & $-1.27$ &    low-alpha &             $2.97$ &                  $2.90$ &                  $3.00$ &    $0.00056$ &         $0.00023$ &         $0.00085$ \\
 170911003601230 & $-0.43$ &   high-alpha &             $2.99$ &                  $2.97$ &                  $3.00$ &    $0.00371$ &         $0.00327$ &         $0.00425$ \\
 170507008301138 & $-0.72$ &   high-alpha &             $2.99$ &                  $2.97$ &                  $3.00$ &    $0.00284$ &         $0.00263$ &         $0.00302$ \\
 140413002701263 & $-0.58$ &   high-alpha &             $2.98$ &                  $2.95$ &                  $3.00$ &    $0.00298$ &         $0.00242$ &         $0.00357$ \\
 161107001601013 & $-0.75$ &   high-alpha &             $2.97$ &                  $2.92$ &                  $3.00$ &    $0.00256$ &         $0.00222$ &         $0.00285$ \\
\hline
\end{tabular}

    \caption{The parameter estimates for the Model C. The mean of the posterior probability distributions ($\alpha_{\rm IMF}$ 
    and $Z_{\rm CC}$) along with 
    the minimum and 
the maximum values that bracket
the 94\% highest density 
interval (HDI) of the posterior 
of each parameter are shown. }
    \label{tab:modelC_results}
\end{table*}

\begin{table*}
    \centering
    \begin{tabular}{lrlrrrrrrrrr}
\hline
    GALAH DR3 ID &  [Fe/H] & OHS subgroup & $\alpha_{\rm IMF}$ & $\alpha_{\rm IMF, min}$ & $\alpha_{\rm IMF, max}$ & $Z_{\rm CC}$ & $Z_{\rm CC, min}$ & $Z_{\rm CC, max}$ & $f_{\rm Ia}$ & $f_{\rm Ia, min}$ & $f_{\rm Ia, max}$ \\
\hline
 160330103301048 & $-1.27$ &    low-alpha &             $2.80$ &                  $2.44$ &                  $3.00$ &    $0.00059$ &         $0.00032$ &         $0.00085$ &       $0.08$ &            $0.06$ &            $0.11$ \\
 170911003601230 & $-0.43$ &   high-alpha &             $0.95$ &                 $-0.99$ &                  $2.28$ &    $0.00565$ &         $0.00540$ &         $0.00580$ &       $0.15$ &            $0.07$ &            $0.23$ \\
 170507008301138 & $-0.72$ &   high-alpha &             $2.68$ &                  $2.18$ &                  $3.00$ &    $0.00299$ &         $0.00294$ &         $0.00302$ &       $0.05$ &            $0.04$ &            $0.07$ \\
 140413002701263 & $-0.58$ &   high-alpha &             $2.44$ &                  $1.62$ &                  $3.00$ &    $0.00398$ &         $0.00371$ &         $0.00415$ &       $0.09$ &            $0.06$ &            $0.14$ \\
 161107001601013 & $-0.75$ &   high-alpha &             $2.51$ &                  $1.68$ &                  $3.00$ &    $0.00277$ &         $0.00261$ &         $0.00285$ &       $0.05$ &            $0.03$ &            $0.09$ \\
\hline
\end{tabular}

    \caption{The parameter estimates for the Model D ($f_{\rm Ch}=0.5$). The mean of the posterior probability distributions ($\alpha_{\rm IMF}$, $Z_{\rm CC}$ and $f_{\rm Ia}$) along with 
    the minimum and 
the maximum values that bracket
the 94\% highest density 
interval (HDI) of the posterior 
of each parameter are shown. }
    \label{tab:modelD_results}
\end{table*}

%%%%%%%%%%%%%%%%%%%%%%%%%%%%%%%%%%%%%%%%%%%%%%%%%%

% Don't change these lines
\bsp	% typesetting comment
\label{lastpage}
\end{document}